\documentclass[11pt]{article}

\newcommand{\lastupdate}{January 22, 2025}
\usepackage[english]{babel}
\usepackage[letterpaper,top=2cm,bottom=2cm,left=3cm,right=3cm,marginparwidth=1.75cm]{geometry}

\usepackage{amsmath}
\usepackage{graphicx}
\usepackage{setspace}
\usepackage[hang, flushmargin]{footmisc}
\usepackage{footnotebackref}
\usepackage{hyperref}

\usepackage[round]{natbib}
\usepackage{fancyhdr,lastpage}%
\usepackage{tikz}
\usepackage{enumerate}
\usepackage{siunitx}
\usepackage{cleveref}

\usepackage{MnSymbol}%
\usepackage{wasysym}%
\usepackage{pifont}
\usepackage[most]{tcolorbox}
\usepackage{fancybox}
\usepackage[utf8]{inputenc}
\usepackage{ifsym}

\usepackage{listings}

\usepackage[notocbasic]{nomencl}
\usepackage{multicol}

\usepackage{tabularx,colortbl}
\usepackage{lscape}

\usepackage{algorithm2e}
\usepackage{verbatim}

\usepackage{arydshln}
\usepackage{pifont}  
\usepackage{enumitem}
\usepackage{har2nat}
\usepackage{csquotes}
\usepackage{eurosym}

\usepackage[normalem]{ulem} 
\usepackage{booktabs}
\usepackage{makecell} 
\usepackage[toc,page]{appendix}

\usepackage{threeparttable}

\definecolor{airforceblue}{rgb}{0.36, 0.54, 0.66}
\definecolor{amethyst}{rgb}{0.6, 0.4, 0.8}
\definecolor{azure}{rgb}{0.0, 0.5, 1.0}
\definecolor{armygreen}{rgb}{0.29, 0.33, 0.13}
\hypersetup{
    colorlinks=true,
    linkcolor=armygreen, 
    urlcolor=armygreen,   
    citecolor=armygreen
}

\usepackage{etoolbox}

\usepackage{float}

\makenomenclature
\renewcommand\nompreamble{\begin{multicols}{2}}
\renewcommand\nompostamble{\end{multicols}}

\fancypagestyle{plain}{
  \fancyhf{}
  \fancyhead[L]{The Journey to Trustworthy AI}
  \fancyhead[R]{Nasr-Azadani and Chatelain}
  \fancyfoot[L]{\small Copyright \copyright \hspace{0.02in} 2024{} Mohamad Nasr-Azadani \& Jean-Luc Chatelain. All Rights Reserved.
  }
  \fancyfoot[R]{\thepage\  / \pageref{LastPage}}
}
\fancypagestyle{firstpage} {
  \fancyhf{}
  \rhead{\textit{Last updated: \lastupdate \\First published: March 8, 2024}} 
  
  \fancyfoot[L]{\small Copyright \copyright \hspace{0.02in} 2024{}
  Mohamad Nasr-Azadani \& Jean-Luc Chatelain. All Rights Reserved.}
  \fancyfoot[R]{\thepage\  / \pageref{LastPage}}
}
\newtcolorbox{fequation}[1][enhanced jigsaw, drop fuzzy shadow, arc=0.5em, top=2mm, bottom=2mm]{ams equation,size=small,#1}

\newtcbox{\bbf}{enhanced jigsaw, drop fuzzy shadow,
    nobeforeafter, notitle,
    boxrule=1pt,arc=1pt,
    left=3pt,right=3pt,top=3pt,bottom=3pt, 
    tcbox raise base,
    fontupper=\sffamily\bfseries}

\crefname{section}{§}{§§}
\Crefname{section}{§}{§§}

\definecolor{Gray}{gray}{0.85}

\newcommand{\topquote}[2][]{%
  \blockquote[#1]{\itshape\sffamily #2}%
  \vspace{\baselineskip} 
  \par\noindent
}

\newcommand{\newabbr}[2]{%
  \nomenclature{\textbf{#1}}{#2}%
  `#2' (#1)%
}

\newcommand{\KK}{Known \textit{Known}}
\newcommand{\KU}{Known \textit{Unknown}}
\newcommand{\UK}{Unknown \textit{Known}}
\newcommand{\UU}{Unknown \textit{Unknown}}

\newtcolorbox{keytakeaways}{
    colback=gray!10, 
    colframe=black, 
    boxrule=0.5pt, 
    arc=5mm, 
    left=5pt, right=5pt, top=8pt, bottom=8pt, 
    before upper={\textbf{Key Takeaways}\par\medskip}, 
    before skip=10pt, 
    after skip=10pt, 
    fontupper=\footnotesize, 
    halign=justify, 
    enhanced, 
    drop shadow=gray!100, 
}

\newtcolorbox{myDisclaimer}{
    colback=gray!10, 
    colframe=black, 
    boxrule=0.5pt, 
    arc=5mm, 
    left=5pt, right=5pt, top=8pt, bottom=8pt, 
    before upper={\textbf{Important:}\par\medskip}, 
    before skip=10pt, 
    after skip=10pt, 
    fontupper=\footnotesize, 
    halign=justify, 
    enhanced, 
    drop shadow=gray!100, 
}

\newcommand{\keypoint}[1]{\begin{itemize}[label=\ding{112}]\item #1\end{itemize}}

\usepackage{caption} 
\definecolor{lightgray}{RGB}{211, 211, 211}

\newcommand{\warning}{\fontencoding{U}\fontfamily{futs}\selectfont\char 49\relax}

\newcommand{\updateOnWHEO}{
\textbf{Update- \textit{\lastupdate}:} The material presented on AI regulations in the USA (in \cref{sec:AI-regulation-by-countries}) is primarily based on Presidential Executive Order 14110, titled `\textit{Safe, Secure, and Trustworthy Development and Use of Artificial Intelligence}'. Unfortunately, as of January 20, 2025, this order has been revoked by the 47th President of the United States. Consequently, the regulatory context discussed no longer reflects the current policies of the Executive Branch of the US Government. We will update the content of our work accordingly in the future.}

\title{\textbf{The Journey to Trustworthy AI:}\\Pursuit of Pragmatic Frameworks}

\author{Mohamad M. Nasr-Azadani\thanks{Email: mohamad@impartialZ.com (Corresponding Author)}\ \quad 
and \quad Jean-Luc Chatelain\thanks{Email: jlc@veraxcap.com}}

\date{} 

\begin{document}
\maketitle

\thispagestyle{firstpage}
\begin{abstract}
This paper, the first installment in a series on \textbf{Trustworthy Artificial Intelligence (TAI)}, reviews various definitions of TAI-- and its \textit{extended family}. Considering the \mbox{\textbf{principles}} respected in any society, TAI is often characterized through a range of attributes or subjective concepts, some of which have led to confusion in regulatory and engineering contexts. We argue against the use of terms such as \textit{Responsible} or \textit{Ethical} AI as substitutes for TAI. And to help clarify any confusion, we suggest leaving them all behind. Given the subjectivity and complexity inherent in TAI, developing a universal framework is deemed infeasible. Instead, we advocate for any approach centered around addressing key attributes and properties such as, \textit{fairness}, \textit{bias}, \textit{risk}, \textit{security}, \textit{explainability}, and \textit{reliability}. We examine the ongoing regulatory landscape, focusing on initiatives in the European Union, China, and the USA, and recognize that geopolitically driven variations in regulating AI pose an additional challenge for multinational companies. We identify \textit{risk} as a core principle in AI regulation and TAI. For example, as outlined in the \textbf{EU-AI Act}, organizations must gauge the \textit{risk level} of their AI products and act accordingly-- or risk paying hefty fines.\\
\\
We compare common modalities of TAI implementation and how multiple cross-functional teams are engaged in the end-to-end process of TAI for any organization. Thus, a brute force approach for enacting TAI renders efficiency and agility, moot. To address this, we introduce our framework `\textit{\textbf{S}et--\textbf{F}ormalize--\textbf{M}easure--\textbf{A}ct}' (\textbf{SFMA}). Our solution highlights the importance of transforming TAI-aware metrics, drivers of TAI, stakeholders, and business/legal requirements into actual benchmarks or tests. Finally, over-regulation driven by panic of powerful AI models can, in fact, harm TAI too. 
Based on GitHub user-activity data, in 2023, AI open-source projects \textit{rose to top projects} by contributor account. Enabling innovation in AI and TAI hinges on independent contributions of the open-source community.
\end{abstract}

\tableofcontents
\thispagestyle{plain}

\section*{Summary Points}
\label{sec:executive-summary}

\begin{keytakeaways}
    \keypoint{Trustworthy AI (TAI) is an evolving concept.}
    \keypoint{There is no `\textit{one-size-fits-all}' solution for TAI.}
    \keypoint{AI will have impacted human civilization at scales not fully understood yet.}
    \keypoint{Meanwhile, there is no need to \textit{panic} or underestimate the impact of AI.}
    \keypoint{THE viable path towards TAI would involve collaboration among social communities, regulators, organizations developing standards, the private sector, open-source communities, academia, and legal scholars-- to name a few.}
    \keypoint{Open-source Software movement has been fueling innovation for decades. Rather than imposing inhibiting restrictions, let's foster it towards advancement of TAI tools and innovations.}
    \keypoint{Experts across various disciplines can play a key role in translating principles of TAI into \textit{attributes} such as safety, reliability, fairness, explainability, etc.}
    \keypoint{There is no single universal framework that can deliver TAI in any organization. Instead, we suggest communities focus on defining and measurement of relevant metrics for various TAI attribute.}
    \keypoint{Several regulatory bodies such as the European Union has approach TAI from a \textit{risk} management perspective.} 
    \keypoint{Clear understanding of uncertainties in AI model's life-cycle should be mapped to risk management frameworks such as the Rumsfeld Risk Matrix (\textbf{RMM}). This enables decision-makers with tools to face and plan for uncertainty.}
    \keypoint{Terms such as `fairness', `bias', `accountability', and `ethical' are \textit{loaded} concepts with roots deeply ingrained in every community's culture, history, societal values, and governance.}
    \keypoint{Association of these terms as `principles' of TAI is context-dependent and, therefore, requires careful `infusion' into any regulatory or engineering system.}
    \keypoint{Mathematically speaking, it has been demonstrated that it is infeasible to satisfy all aspects of AI fairness concurrently.}
    \keypoint{Therefore, discussions surrounding fair AI and required policies can turn subjective and philosophical, e.g. `least harmful path' \emph{vs} `most profitable path'.}
\end{keytakeaways}

\section*{Disclaimer}
\label{sec:disclaimner}

\begin{enumerate}

    \item[\warning] \updateOnWHEO

    \item[\warning] While discussions in this work aim for longevity, AI regulations and legislation are still evolving in many countries. Therefore, some discussions may require updates as new regulations emerge.
    
    \item[\ding{36}] In this work, we do not consider the following AI-systems: 
    \begin{enumerate}[label=\ding{56}]
        \item \textbf{AI-controlled and fully autonomous robotic systems:} For a recent survey, cf. \citet{ingrand2017deliberation,kunze2018artificial}.
        \item \textbf{In vivo AI-powered synthetic biology and biotechnology:} For example, \textit{Xenobots} (cf. \citet{blackiston2021cellular} and \citet{kriegman2021kinematic}).
        \item \textbf{Self-evolving and self-replicating AI:} For example, cf. \citet{AISelfreplication}.
        \item \textbf{Quantum machine learning:} For a recent survey, cf. \citet{zhang2020recent}.
    \end{enumerate}
\end{enumerate}

\section{Context}
\label{sec:context}

Across the globe, many governments and legislative bodies are actively working to regulate the development and use of \newabbr{AI}{Artificial Intelligence}, cf. \citet{reutersAITracker2023}. For instance, President Biden's recent executive order on `safe, secure, and trustworthy AI' issued in October 2023 \{\textbf{Update \textit{\lastupdate}: Unfortunately, this executive order has been revoked by the 47th President of the United States}\} \footnote{\textbf{Update \textit{\lastupdate}:} Presidential Executive Order (EO)-14110 was first released by the White House on October 30th, 2023 and unfortunately later revoked by the 47th President of the United States on January 20th, 2025. The full draft can be obtained here \citet{BidenAIEO2023} or \citet{BidenAIEO2023UCSB}.} was quickly followed by a similar announcement from the \newabbr{EU}{European Union} in which the EU members unanimously reached a \textit{political agreement} to regulate AI\footnote{Commonly known as the `\textbf{EU-AI Act}', this legislation is expected to go into effect in 2025 or 2026 and has been hailed by many as the first comprehensive legislation for TAI.} (\citet{europarlAIAct}).

The primary impetus behind the ongoing regulation of AI is the wide-ranging impact it will have on every facet of human life. In essence, AI in conjunction with existing/emerging technologies such as \newabbr{IoT}{Internet of Things}, 5G/6G\footnote{6G telecommunication network paradigms are still in the research stage, cf. \citet{jiang2021road}. Generally speaking, 6G's mission is to build the communication platform which a hybrid world consisting of physical and digital realities, e.g. \newabbr{AR}{Augmented Reality}, can function, seamlessly. Commercial 6G is expected to arrive in late 2020s or early 2030s, cf. \citet{Ericsson6G}.}, and \newabbr{DX}{Digital transformation} is poised to impact the so-called \newabbr{4IR}{Fourth Industrial Revolution}, cf. \citet{philbeck2018fourth, french20214th}.

In general, \textit{risk} and \textit{uncertainty} are considered \textit{intrinsic properties} in many autonomous systems. AI-powered systems are not exempt from this classification either. Hence, the term \newabbr{TAI}{Trustworthy Artificial Intelligence} has been coined, representing multi-disciplinary research areas tackling the `\textit{distrust}' in AI systems. 
With the remarkable performance of recent AI products, such as ChatGPT, regulatory bodies have accelerated their efforts to pass legislation. While we recognize substantiated concerns raised by public and prominent research scholars\footnote{
Recently, in response to remarkable \textit{human-like} capabilities demonstrated by a new family of AI models called \newabbr{GPT}{Generative Pre-trained Transformer} and \newabbr{LLM}{Large Language Model}, public opinion along with that of prominent AI academics such as \href{https://scholar.google.com/citations?hl=en&user=JicYPdAAAAAJ}{Professor Geoffrey Hinton}, has raised serious concerns about the potential existential threat posed by AI models, e.g. \citet{barrat2023our}. While we do not discount the possibility of \textit{doomsday events} triggered by `AI-\textit{gone-rogue}', addressing circumstances that could lead to catastrophes of such magnitude is beyond the scope of this article. For more on this, we refer the reader to recent surveys, cf. \citet{galanos2019exploring, carlsmith2022power, bucknall2022current, federspiel2023threats}.}, we caution against the over-regulation of AI. In several cases, open-source community is being targeted which could hinder-- the much needed-- innovation from these vibrant communities to enable TAI. We will further discuss this topic in \cref{ssec:where-is-ai-header-github}. 

Considering efforts to regulate AI, we argue that TAI-- along with the disciplines surrounding it--has played a unique role in the path ahead: It has motivated cross-functional collaboration among experts and stakeholders, and regulatory entities to: 
\begin{enumerate}[label=\ding{43}]
\item Understand cutting-edge AI technologies,
\item Assess the near- and long-term impact of AI on society and the economy,
\item Propose new policies, standards, and frameworks,
\item Solicit and incorporate feedback from the public domain in TAI policies,
\item Enact new or revise existing laws, standards, and guidelines. 
\end{enumerate}

\noindent
In this work, we hope to provide our point of view on `\textit{journey}' towards the realization of TAI. 
In doing so, in part 1, we demonstrate how numerous `principles' of TAI\footnote{So far, several domestic and international organizations have released lists of `\textit{\textbf{TAI mission.}}'. While there are common items their compiled lists, we emphasize that every organization prioritizes a certain aspect of human life and society that is aligned with its mission when publishing principles of TAI. For example, IEEE is focused on building robust standards for engineering applications. Alternatively, UNESCO is focused on human rights and education. We will discuss these later in the text.} could be aggregated and be ``\textit{transformed}'' into tangible `frameworks' enabling TAI within any organization. 

We provide a summary of characterizations as well as \textit{taxonomy} used by multi-disciplinary scholars addressing TAI and its derivatives, e.g. \newabbr{XAI}{eXplainable Artificial Intelligence} or AI fairness. In \cref{ssec:our-set-formalize-measure-framework-solution} we introduce our proposed solution called `\textbf{\textit{Set, Formalize, Measure, and Act}}', a simple yet powerful framework towards TAI for enterprise.  

In part 2, we provide an overview of recent advancements in statistical and data-driven techniques for quantifying critical metrics representing every dimension of TAI. We will compare different modalities of implementing TAI, prioritizing `Trustworthy-By-Design' frameworks.

\section{Trustworthy AI: Too Many Definitions or Lack Thereof?}
\label{sec:trustworthy-ai-definition}

We argue that there is not concrete definition for the terminology `trustworthy artificial intelligence' insofar as it has been characterized by the desired attributes in a particular discipline such as engineering, education, economy and markets, and public policy, cf. \citet{stix2022artificial}. 
\begin{table}[htbp]
    \centering
    \caption{Attributes extracted from principles defining `Trustworthy AI' that have been announced by various entities. For a complete list of principles for each, see \cref{ssec:apx-guiding-principles-from-entities}.}
    \label{tab:TAI-definitions-entities}
    \resizebox{\textwidth}{!}{ 
    \begin{tabular}{c||c|c:c:c:c:c:c:c:c:c:c} 
        \hline
        \hline
        \thead{\textbf{Entity}\\\textbf{Name}} & \thead{\textbf{Framework or}\\\textbf{Theme}} & \thead{\textbf{Safe}\\\textbf{Secure}} & \thead{\textbf{Privacy-}\\\textbf{enhanced}} & \thead{\textbf{Explainable}\\\textbf{Interpretable}} & \thead{\textbf{Transparent}\\\textbf{Accountable}} & \thead{\textbf{Human}\\\textbf{Oversight}} & \thead{\textbf{Robust}\\\textbf{Resilient}} & \thead{\textbf{Reliable}\\\textbf{Valid}} & \thead{\textbf{Prioritizing}\\\textbf{Humans}} & \thead{\textbf{Fair}} & \thead{\textbf{Literacy}\\\textbf{Awareness}} \\
        \hline
        \cellcolor{gray!25}\textbf{NIST} & \thead{Risk\\Management} & $\checkmark$ & $\checkmark$ & $\checkmark$ & $\checkmark$ & & $\checkmark$ & $\checkmark$ &  & $\checkmark$ & \\
        \hline
        \cellcolor{gray!25}\textbf{UNESCO} & \thead{Human\\Rights} & $\checkmark$ & $\checkmark$ & $\checkmark$ & $\checkmark$ & $\checkmark$ & $\checkmark$ & & $\checkmark$ & $\checkmark$ & $\checkmark$ \\
        \hline
        \cellcolor{gray!25}\textbf{IEEE} & \thead{Trustworthy-\\by-Design} &  &  &  & $\checkmark$ & $\checkmark$ & & & $\checkmark$ & & $\checkmark$ \\
        \hline
        \cellcolor{gray!25}\textbf{OECD} & \thead{Democracy \& \\Market Economy} & $\checkmark$ &  & $\checkmark$ & $\checkmark$ &  & $\checkmark$ &  & $\checkmark$ & $\checkmark$ & \\
        \hline
    \end{tabular}
    } 
\end{table}

\noindent 
As such, any entity that aims to define TAI should consider factors such as applications (or services), business goals, legal context, and parties involved, amongst other important elements (for a recent review on TAI definitions and taxonomy, we refer the reader to \citet{thiebes2021trustworthy,jacovi2021formalizing}).

In the remainder, we recap the principles for TAI recommended by various governmental and other international entities, namely the \newabbr{NIST}{National Institute of Standards and Technology}, the \newabbr{UNESCO}{United Nations Educational, Scientific and Cultural Organization}, the \newabbr{IEEE}{Institute of Electrical and Electronics Engineers}, and the \newabbr{OECD}{Organization for Economic Co-operation and Development}. 

We have selected this diverse list of entities to show the commonalities in TAI principles despite their varying missions. In \cref{tab:TAI-definitions-entities}, we provide an aggregated view to demonstrate how independent international or domestic units focus on various \textit{features} in an AI system to be considered a TAI. For more details on the principles for each entity, see \cref{ssec:apx-guiding-principles-from-entities}.

\subsection{Trustworthy AI: \textit{Attribute} or \textit{Property}?}
\label{ssec:standardization-TAI-attribute-property}

The process of building TAI rapidly turned into an amalgamation of an ever-growing number of attributes expected from any AI system and its output. For example, `\textit{Fairness in AI}', `\textit{AI Safety}', `\textit{Secure AI}', `\textit{Transparent AI}', `\textit{Explainable AI}', `\textit{Interpretable AI}', `\textit{Black-box AI}', `\textit{Responsible AI}', `\textit{Robust AI}', `\textit{Resilient AI}', `\textit{Ethical AI}', `\textit{Reliable AI}', `\textit{Privacy-enhanced AI}', `\textit{Accountable AI}', and `\textit{Federated AI}' are common examples of such attributes. In other words, and out of necessity, we have been ``cooking'' this topic in a \textit{magic pot} with ``chefs'' from various disciplines. 

We must keep in mind that most of the aforementioned terms characterizing TAI do not possess a universally accepted definition. A few terms are used interchangeably. For example, consider `interpretability' and `explainability' that are used synonymously. From an engineering perspective, `explainable AI' and `interpretable AI' point to two distinct technical concepts. For instance, \textit{\textbf{outputs}} returned by a \newabbr{DNN}{Deep Neural Network} model can be \textit{\textbf{explained}} using algorithms such as LIME (\citet{ribeiro2016should}) despite DNNs categorized as not interpretable\footnote{In the literature, DNNs are categorized as `Black-box' AI models. In layman's terms, internal structure of DNNs can be so complex that understanding \textit{\textbf{how}} they produce their outputs can be very hard-- if not impossible.}. In contrast, it is widely accepted that term `interpretability' should be classified as an intrinsic property when selecting a family of AI model. For example, deploying a `Decision Tree Classifier' as an AI product provides `interpretability' almost at no additional computate cost. This is because of its inherent `\textbf{\textit{If-Then-Else}}' topology when computing an outcome. To summarize, every feature (attribute) utilized to characterize TAI is either:
\begin{enumerate}
        \item[I.] \textbf{An Intrinsic Property:} An inherent attribute or a characteristic of an \textit{object} which does not depend on its external environment, relationship, or conditions. Hardness and mass are the intrinsic properties of a diamond. We argue that classifying properties of a TAI system is necessary and can simplify the frameworks, legal ramifications, and implementation techniques. An example of an intrinsic property in an AI product is its degree of ``Black-Boxness''. In this context, `Black-box' AI--a term predominantly used by AI engineers--is an \textit{intrinsic property} of an AI model. It indicates a category of AI model where the underlying mathematical reasoning is non-linear and complex to be readily understood by humans.
        \item[II.] \textbf{An Extrinsic Property:} An extrinsic property on an object or substance depends on the external factors and relationships with other external objects. For instance, temperature of an object depends on the surrounding environment. In an AI system, such properties can be assumed an `add-on' to an existing AI model. For instance, an AI team can make an existing computer vision model `secure' by adding additional layers to it could be deployed in a high-risk use-case such as self-driving cars.  In other words, a company could initially train a sophisticated and reliable computer vision model without implementing `security', and subsequently apply methods to add this (extrinsic) property in an on-demand manner. 
\end{enumerate}

\noindent
Next, one may ask why these categories matter? Without going into details, agreeing on such classification clearly early on could help any organization with implementing and maintaining TAI in its product life-cycle\footnote{
As of today, a universally accepted framework for the AI product life-cycle does not exist, in contrast to the well-established mature software development life-cycle. This absence can be attributed, in part, to the diverse organizational structures, business processes, and modes of AI-model integration within enterprises.}. Consider the common yet important decisions impacting AI product life-cycle 
\begin{enumerate}[label=\ding{43}]
    \item \textbf{Metric Selection:} Map requirements to metrics. For example, there are numerous ways to evaluate `\textit{fairness}' of a loan approval AI model. Selection and compute the `fairness-score' may not be trivial.
    \item \textbf{Enterprise Risk Management (ERM):} Since any data-driven product inherently is not `\textit{bullet-proof}', risk-assessment and management frameworks used currently by an enterprise can impact the \newabbr{UQ}{Uncertainty Quantification} techniques which may be directly tied to ERM.
    \item \textbf{Resource Allocation:} Allocate and plan on resources such as human \newabbr{SME}{Subject Matter Expert}, continuous monitoring and improvement platforms for AI systems running in production.
    \item \textbf{Legal Compliance:} Understanding the risks involved in violating legal obligations is the first step to plan and absorb the inherent risk associated with any AI-product.
\end{enumerate}

\subsection{Challenges-turned-into Myths Surrounding Trustworthy AI}
\label{ssec:mystery-mist-misconstrued}

It is fascinating to watch how the topic of Trustworthy AI-- and its variants-- has been debated by scholars and policy makers across a wide number of domains. Several scholars argue that assigning terms such as `trustworthy' or `responsible' to AI (in the context of legislation) may confuse various sectors. If not properly differentiated, it ultimately undermines proper implementation and enforcement of TAI, cf. \citet{freiman2023making,laux2024trustworthy}.

\subsubsection{Example Myths about Trustworthy AI}
\label{ssec:tai-myths}
In order to bring clarity surrounding TAI and its definitions, it is important to, first, recognize questions or assumptions that eventually rendered technical challenges as \textit{myths}. Here are a few examples:
\begin{enumerate}[label=\ding{56}]
    \item \textbf{Myth:} Products using AI are autonomous; therefore, their ``\textit{decisions}'' cannot be comprehended or defended.
    \item \textbf{Myth:} We (humans) are not capable of rationalizing the decisions made by black-box AI models.
    \item \textbf{Myth:} We cannot ``\textit{control}'' the decisions of an AI system.
    \item \textbf{Myth:} The only \textit{reason} for an AI model to act \textbf{\textit{unethically}} is due to its training performed by a human (or a human-supervised system). 
    \item \textbf{Myth:} Any AI model that is trained on real-world data-- echoing human history, values, and the evolution of society-- cannot have its harmful biases mitigated.
    \item \textbf{Myth:} Any decision made solely based on \textit{human intuition} always outperforms than that of an AI system (or \emph{vice versa}). 
    \item \textbf{Myth:} With the emergence of larger and more powerful AI models, e.g. ChatGPT, humans are to be completely removed from the decision-making process.\footnote{Currently, it is widely accepted that the human brain outperforms the `\textit{best}' \newabbr{AGI}{Artificial General Intelligence} system. Qualities such as `\textit{out-of-the-box}' thinking, and `\textit{causal reasoning}' (\citet{,bishop2021artificial}) are considered example `super-powers' of human brain. There is a strong consensus in the scientific community that by only increasing the \textit{\textbf{size}} and enhancing the \textit{\textbf{capacity}} of AI models, we cannot produce AGIs capable of outsmarting humans in every capability, cf. \citet{Lecun2023,fjelland2020general}.}
\end{enumerate}

\noindent
While having healthy debates around these topics or myths is always welcome, it should not promote valid concerns into paralyzing or panic shutting down progress in AI. 

\subsection{Trusting AI Systems: A Complicated Relationship with Humans}
\topquote[Antoine Rivarol (1753--1801); A French writer]{In republics, the people give their favor, never their trust.}

One might simply ask: `\textbf{What is trust}?'. To make matters more complicated, there is no unified definition for `\textbf{\textit{trust}}' across different disciplines. Psychologists consider trust a \textbf{cognitive attribute of the human mind}\footnote{\Citet{rotter1980interpersonal} defines trust as: `\textit{Cognitive learning process obtained from social experiences based on the consequences of trusting behaviors}'.}, sociologists associate trust with \textbf{human relationships}\footnote{As in sociology, trust is defined as (\citet{gambetta2000can}): `\textit{Subjective probability that another party will perform an action that will not hurt my interest under uncertainty and ignorance'.}}, and economists argue that trust\footnote{In \citet{james2002trust} and in the context of economic systems, trust is defined as: `\textit{Expectation upon a risky action under uncertainty and ignorance based on the calculated incentives for the action}'.} can, in fact, be `\textbf{calculated}' (\citet{granovetter2018economic}). For a comprehensive list of definitions for \textit{trust} across various disciplines, we refer reader to \cite{cho2015survey} and references therein. 

The presence and influence of decisions made by automatic algorithmic systems is undeniable. Recently, terms such as `\textit{algocracy}' (algorithmic government) have been used to describe potential `futuristic' governments. Such ideas are not far-fetched. For example, a software named COMPAS is used in justice systems in the USA to help judges assess the likelihood that a defendant becomes a recidivist (we discuss this in \cref{ssec:fairness-compas-example}). Additionally, it is estimated that the majority of trading performed on Wall Street is carried out by autonomous algorithms and trading bots, cf. \citet{patterson2013dark,menkveld2016economics,isidore2018machines}.

\subsubsection{How do we (Humans) Trust the \textit{Unknown}: It is always a Process}

As history has shown us, when faced with new technologies such as \textit{electricity}, \textit{the Microwave oven}, or AI, many people typically respond with \textit{justified} skepticism, resistance\footnote{An example is the \textit{Printing Press} introduced in the 15th century, which faced resistance from Catholic Church as well as monarchies in Europe. Such entities relied on censorship, manipulated licensing systems, and enforced heavy penalties for `unapproved printing' to limit the impact of the printing press on educating people. With education becoming more accessible to a wider audience, the control of religious rulers, governments, and monarchs over the people was jeopardized, cf. \citet{pardue2012printing,robertson2015censorship}.}, fear, and sometimes, complete backlash against innovations like `Google Glass' (\citet{kudina2019ethics}). 

To overcome such resistance, it is important to harness the power of `\textit{trust}'. The successful interplay of `Realizing Trust' and `Human Societies' commonly undergoes several steps (cf. \citet{frischmann2018re,lankton2015technology}) listed below:

\begin{enumerate}[label=\alph*)]
    \item \textbf{Establishing Trust:} Properly and transparently `introduce' new technology to the community. In addition, `educate' users on how to interact and utilize it. 
    \item \textbf{Building Trust:} Allow users interact with the new technology in a safe and guided manner. When many users \textit{consistently} have positive experience in their engagement with the new system and notice that the outcomes align with their `\textit{ethical}' norms, it can be assumed that \textit{trust is built}. 
    \item \textbf{Maintaining Trust:} Requires ongoing effort to ensure continuous improvement, demonstrating willingness and adaptability to evolving challenges, and open and honest communication channels with their users.
    \item \textbf{Rebuilding Trust (If needed):} As no system is perfect, when a new system fails, restore and rebuilding trust requires steps to remediate the problem, remove any culprit(s), and be transparent with its users upon completion of conducted \newabbr{RCA}{Root Cause Analysis}.
    \item \textbf{Sustaining Trust:} Requires steps to encourage the involvement of communities in the long-term engagements and fostering the technology at hand by providing feedback channels and a focus on long term value.
\end{enumerate}

\noindent 
Without delving into specifics, we note that the process of building trust between `an individual person' (as opposed to a group or a community) and a new technology can differ from steps discussed above. Psychological and biological variations could significantly influence the outcome. 

\subsubsection{UK Home Office's Biased Algorithm: An Example of Failure in Building Trust}

The UK Home Office faced criticism for its use of an AI algorithmic system in processing visa applications, which came to light in 2018, cf. \citet{gualdi2021artificial}. Before it was publicly labeled as a biased and `racist algorithm' \citet{bbc2020VisaRacistAlg}, this AI engine had been built to ``\textit{streamline}'' the heavily backlogged visa application process. Towards this, given a visa applicant, this AI product ``categorized'' applications into various risk levels and identified ``high-risk'' cases for further scrutiny.

Let's recap the challenges and how actions (or lack thereof) damaged `Trust' between immigrant communities and the \href{https://www.gov.uk/government/organisations/home-office}{UK Home office}: 

\begin{enumerate}[label=(\alph*)]
    \item \textbf{Familiarity and Consistency:} The introduction of the algorithm disrupted the familiarity for visa applicants, as `black-box' and automated system suddenly played a significant role in the decision-making process.

    \item \textbf{Transparency:} The new algorithm lacked  \textit{transparency} in its decision-making process. 
    \begin{quote}
    ``\textit{Potentially life-changing decisions are partly made by a computer program that nobody on the outside was permitted to see or to test}'', Cori Crider, Foxglove (\citet{cnet2023}).
    \end{quote}
    Visa applicants were not informed about the criteria and factors used by this algorithm which determined their ``\textit{risk level}'', leading to concerns about accountability of the system.

    \item \textbf{Perceived Competence:} Concerns about the origin of the new algorithm (and its training dataset), accuracy, and fairness risk scores raised questions about the competence of the UK Home Office in overseeing and implementing its new visa processing system using AI.
    \begin{quote}
``\textit{... Researchers from \href{https://www.foxglove.org.uk/}{Foxglove} and the \href{https://www.jcwi.org.uk/}{JCWI} believed it was built in house by the \{UK\} government rather than brought in from a private company. They allege that the government is being purposefully opaque about the algorithm because it discriminates based on the nationality of the applicant, and that it doesn't want to release a list of the countries it considers high risk into the public domain.}'' (\citet{cnet2023})
    \end{quote}
    
\item \textbf{User Control:}
   Visa applicants (or independent legal entities) had limited ``control'' (if any) over the decision-making process. The lack of transparency did not allow them to address issues or to petition a decision made by the UK Home Office in a meaningful manner-- in an event of a rejection outcome. 
   
\item \textbf{Long-Term Relationship Building:}
   Trust issues stemming from the opacity (lack thereof) of the algorithmic decision-making process potentially harmed the long-term relationship between the government and visa applicants. 
\begin{quote}   
``\textit{We also discovered that the algorithm suffered from `feedback loop' problems known to plague many such automated systems - where past bias and discrimination, fed into a computer program, reinforce future bias and discrimination.  Researchers documented this issue with predictive policing systems in the US, and we realised the same problem had crept in here.}'' (\citet{foxglove2020})
\end{quote}
Given the circumstances, rebuilding trust requires addressing concerns from all parties involved, increasing transparency in existing or future models, and providing avenues public-facing auditing mechanisms.

\item \textbf{Community Involvement:}
   This automated system--which was in place for five years-- incorrectly rejected numerous visa applications to the UK based solely on the \textit{applicant's country of origin}. This could have been remediated earlier if immigration advocacy groups, independent technical firms, legal councils, and applicants, were all included in discussions and oversight about the use of automated decision-making tools. 
\end{enumerate}

\section{Complexities and Challenges}
\label{sec:challenges}
\begin{keytakeaways}
    \keypoint{Let's avoid a `wild-goose chase': AI is not the ``\textit{responsible}'' agent in the room: Its users and companies are.}    
\end{keytakeaways}

In this section, we aim to provide our insight on why there is no `one-size-fits-all' solution for TAI. 

\subsection{``\textit{Responsible AI}'': A Confusing Term that should be Left Behind}
\label{ssec:responsible-ai-confusing-term}

It is safe to assume that by now, AI and related disciplines such as \newabbr{ML}{Machine Learning} or Data Science, are independent scientific paradigms, akin to mathematics or statistics. Just as no one expects to comprehend phrases such as `Responsible Mathematics', we argue that the term `Responsible AI' is meaningless. Since AI has already become an essential tool for aiding product teams, it is actors how decide how to utilize it in their business. Simply put, `Irresponsible actors/engineers/managers' do exist, not so much `Responsible AI' and its evil twin, `Irresponsible AI'. 

\subsubsection{Mathematics cannot be Held ``Responsible'', nor should AI}
\label{sssec:math-cannot-be-held-responsible}
Mathematics, inherently, cannot be held accountable; rather, the responsibility lies with the individual or entity utilizing mathematics. In a similar vein, same principles apply to other scientific disciplines including AI. Without digging into current legal and philosophical debates surrounding agency as well as accountability surrounding any autonomous system, in scenarios where an AI-product `\textit{is}' in charge of making decisions autonomously and independently, entity who passed on such responsibility to this product would be held liable. 

\subsubsection{Examples: Achieving Clarity by not Expecting a Product to be ``Responsible''}
\label{sssec:example-dont-use-responsible-ai}

To drive our point home, let's imagine we encounter news headlines such as the following list:
\begin{enumerate}[label=\ding{56}]
    \item MagicKar, a car manufacturing company, is making a `\textit{Responsible Self-driving Car}' as their next model. 
    \item President of University of MarsY forms a committee to develop a framework for `\textit{Responsible Computer Science}'.
    \item An online search engine company, called Tix-Tax-Tox, announces the release of its new `\textit{Responsible Search Engine}'.
\end{enumerate}

\noindent 
Statements above while \textit{grammatically} correct, are not semantically comprehensible-- to say the least. Any organization tasked to build a `responsible product X' will have follow-up questions such as `\textit{a) What is considered a responsible car? or b) Is this a legal or ethical mandate?}'. 
In response to such clarifying questions, a person has to only use context-aware and relevant terms to describe `being responsible or `acting responsible':
\begin{enumerate}[label=\ding{52}]
    \item ..., MagicKar, is making a `\textit{\sout{Responsible} \textbf{Safe} Self-driving Car}' as their next model. 
    \item ..., a committee to develop a framework for `\textit{\sout{Responsible} \textbf{Transparent \& Resilient} use of Computer Science}'.
    \item ..., company announces the release of its new `\textit{\sout{Responsible} \textbf{Unbiased} Search Engine}'.
\end{enumerate}

\subsubsection{An Undesirable Outcome for AI Industry: ``\textit{Responsibility-as-a-Service}''}
\label{sssec:respons-as-a-service}

The title says it all... Considering the complexities of TAI and soon-to-be-enacted AI regulations, this scenario may occur seamlessly-- if not already. 
Assuming `responsible' as a characteristic of an AI system can marginalize the significant effort, technical debt, legal considerations, and human expertise required. Driven by a highly competitive market in AI, we observe signs of such shift in building large scale AI-enabled products: In essence, a company first trains an AI model only focusing on its \textbf{performance} and \textbf{accuracy} geared towards business outcome. Once this model is trained, company attempts to find out what and how it can \textbf{make} it `\mbox{\textit{responsible}}' without deteriorating now-trained AI model's accuracy-- as long as the upgraded AI model performance somehow remains within the legal bounds. If such bounds are relaxed or turned more stringent, this company would only expand or shrink their team or resources allocated for ``RaaS-AI'', accordingly. 

We hope that we have convinced you that the term `Responsible AI' is not a suitable ambassador for TAI, hitherto. This is further pronounced in legislation and regulatory contexts. Practically speaking, proper integration and usage of AI models in any product, application, or services approved by governing bodies, ought to be carried out following a multi-tiered legislative or regulatory enforcement. Many countries have recently started experimenting multi-tiered regulation of AI. Some even have set \textit{risk} as the core element in their TAI regulatory frameworks. We discuss this further in \cref{sec:AI-regulation-by-countries}.

\subsection{Trust and the Parties Involved}
\label{ssec:parties-involved}

\begin{figure}
    \centering
    \includegraphics[width=0.70\linewidth]{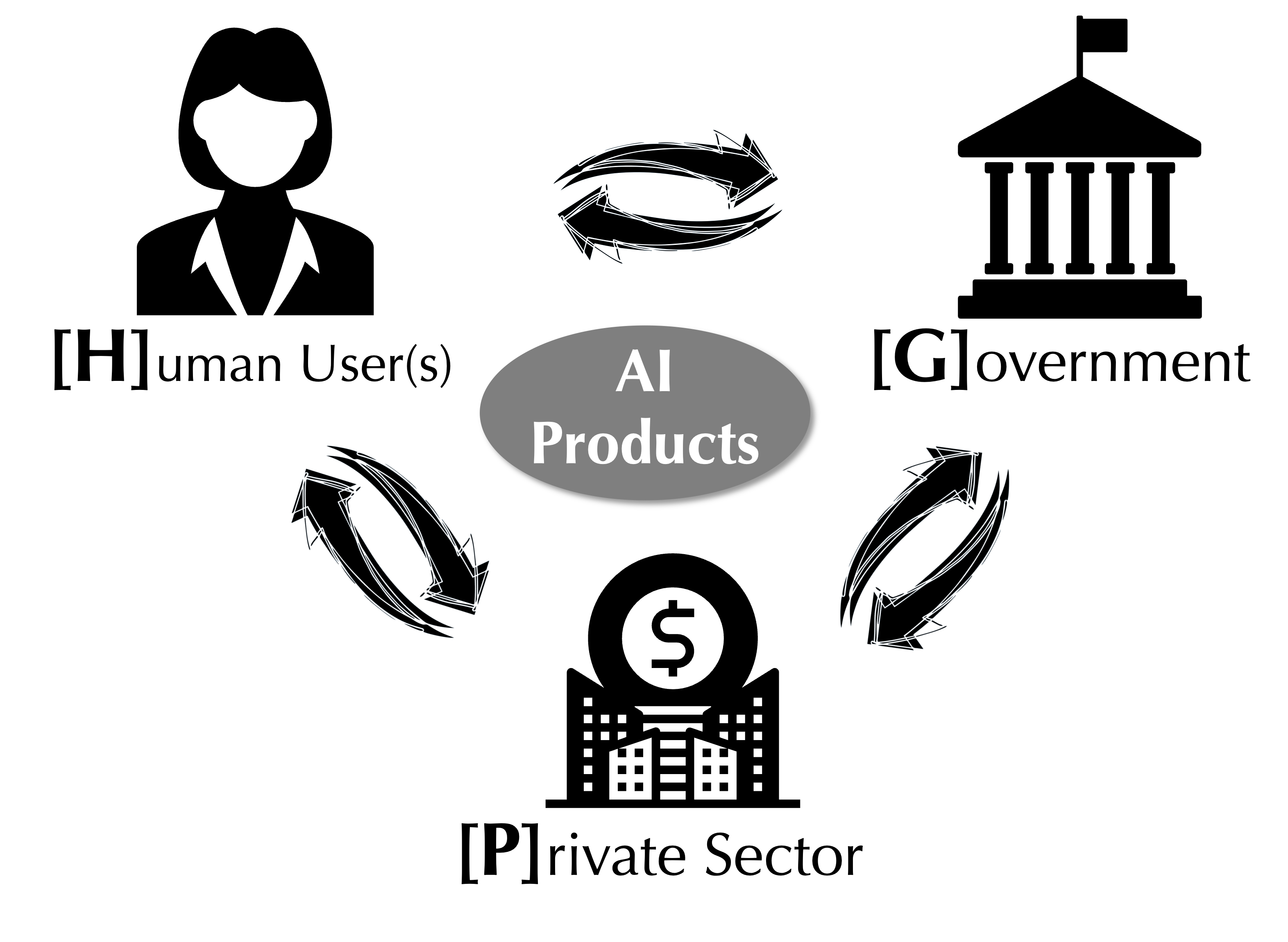}
    \caption{Main parties involved in assessing `Trustworthy AI' in a product or service; \textbf{H}uman end user (or community); \textbf{G}overnment; and the \textbf{P}rivate sector. Note that for every two entities, any acceptable TAI framework should be equipped to address the any professional (two-way) interactions.}
    \label{fig:human-government-private}
\end{figure}

Figure \ref{fig:human-government-private} shows three distinctive entity type that can interact in a business or professional context. In essence, either one- or two-way interactions\footnote{Note that there are other possible categories, e.g. self-self and three-way interactions. For the sake of simplicity, we do not discuss them here.}, need be considered when a target TAI framework is to be developed.

\noindent
Below we categorize varieties of `two-way interactions' (see \cref{fig:human-government-private}) that can occur in any professional or social context:
\begin{itemize}
    \item $\colorbox{lightgray}{\(\textrm{H}\ \leftrightarrows\ \textrm{G}\):}$ Human interactions with Government (and \emph{vice versa}). Example: Use of AI by judiciary system and the rights of citizens.  
    \item $\colorbox{lightgray}{\(\textrm{H}\ \leftrightarrows\ \textrm{P}\):}$ Human interactions with Private entities (and \emph{vice versa}). Example: Use of AI by a bank to approve/reject a citizen's loan application. 
    \item $\colorbox{lightgray}{\(\textrm{G}\ \leftrightarrows\ \textrm{P}\):}$ Government interactions with private entities (and \emph{vice versa}). Example: Use of AI by \newabbr{FTC}{Federal Trade Commission} to investigate reports of illegal activities carried out by a specific bank.
    \item $\colorbox{lightgray}{\(\textrm{G}\ \leftrightarrows\ \textrm{G}'\):}$ One government entity interacting with another government entity (and \emph{vice versa}). Example: The Supreme Court of USA investigating data-backed claims regarding `gerrymandering' in a particular state.
    \item $\colorbox{lightgray}{\(\textrm{H}\ \leftrightarrows\ \textrm{H}'\):}$ Human interacting with another human. Example: A citizen using AI to publish fake images of a former colleague.
    \item $\colorbox{lightgray}{\(\textrm{P}\ \leftrightarrows\ \textrm{P}'\):}$ Private entity interacting with another private entity. Example: An internet search engine giant throttling internet speed only for iPhone (as opposed to Android) users.
\end{itemize}

\noindent 
Attributes associated with TAI are directly or indirectly be impacted by the family of interaction and entities involved. For example, explainability-- a pillar in any TAI framework-- requirements are different for a government's legal investigation \emph{vs} a social media user requesting explainability on how her activity data was used to see particular advertisements. 

\subsection{Geographical and Geopolitical Considerations}
\label{ssec:trustai-geography}

\begin{figure}
    \centering
    \includegraphics[width=0.99\linewidth]{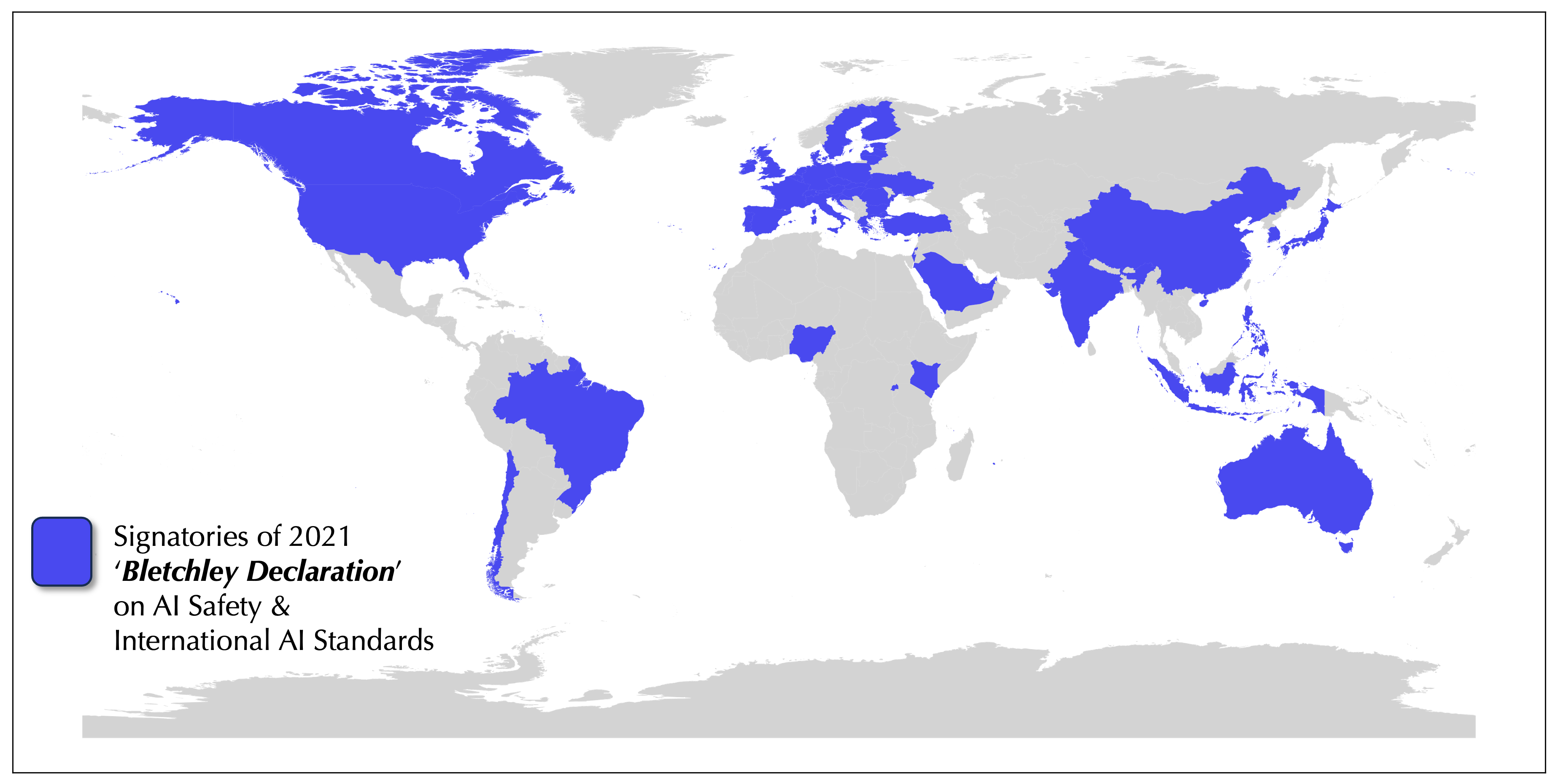}
    \caption{The first international summit on AI Safety held in November 2023 in Bletchley, UK. Twenty eight countries signed `Bletchley Declaration'. List of countries retrieved from \cite{toney2023who}.}
    \label{fig:world-bletchley-declaration}
\end{figure}

The first international conference called `AI Safety Summit' was held in the United Kingdom in November 2023. This event concluded with 28 countries signing an agreement known as the `\textbf{Bletchley Declaration}' (see \cref{fig:world-bletchley-declaration}). First of its kind, Bletchley Declaration focuses on the challenges and risks of AI and, therefore, seeks cooperation among international communities and countries to establish cooperating channels to mitigate risks posed by AI (\citet{bletchleydeclaration2023}). While Bletchley Declaration is a good example of international cooperation to regulate AI, geopolitical dynamics play an important role in making or breaking such efforts.

\noindent
Consider leading global powers such as the United States, China, and the European Union. The United States, with its tech giants and well-established innovation ecosystems, sets critical trends in the development of AI and TAI centered on markets, while China's state-driven approach where it prioritizes a centralized authority to regulating AI in areas such as \textit{content generation} or \textit{recommendation systems}. The EU, however, following its existing strict data privacy and ethical standards such as GDPR, is now taking a strict approach to regulate AI through with `risk' at its core (we discuss this in \cref{sec:AI-regulation-by-countries}).

\begin{figure}
    \centering
    \includegraphics[width=0.99\textwidth]{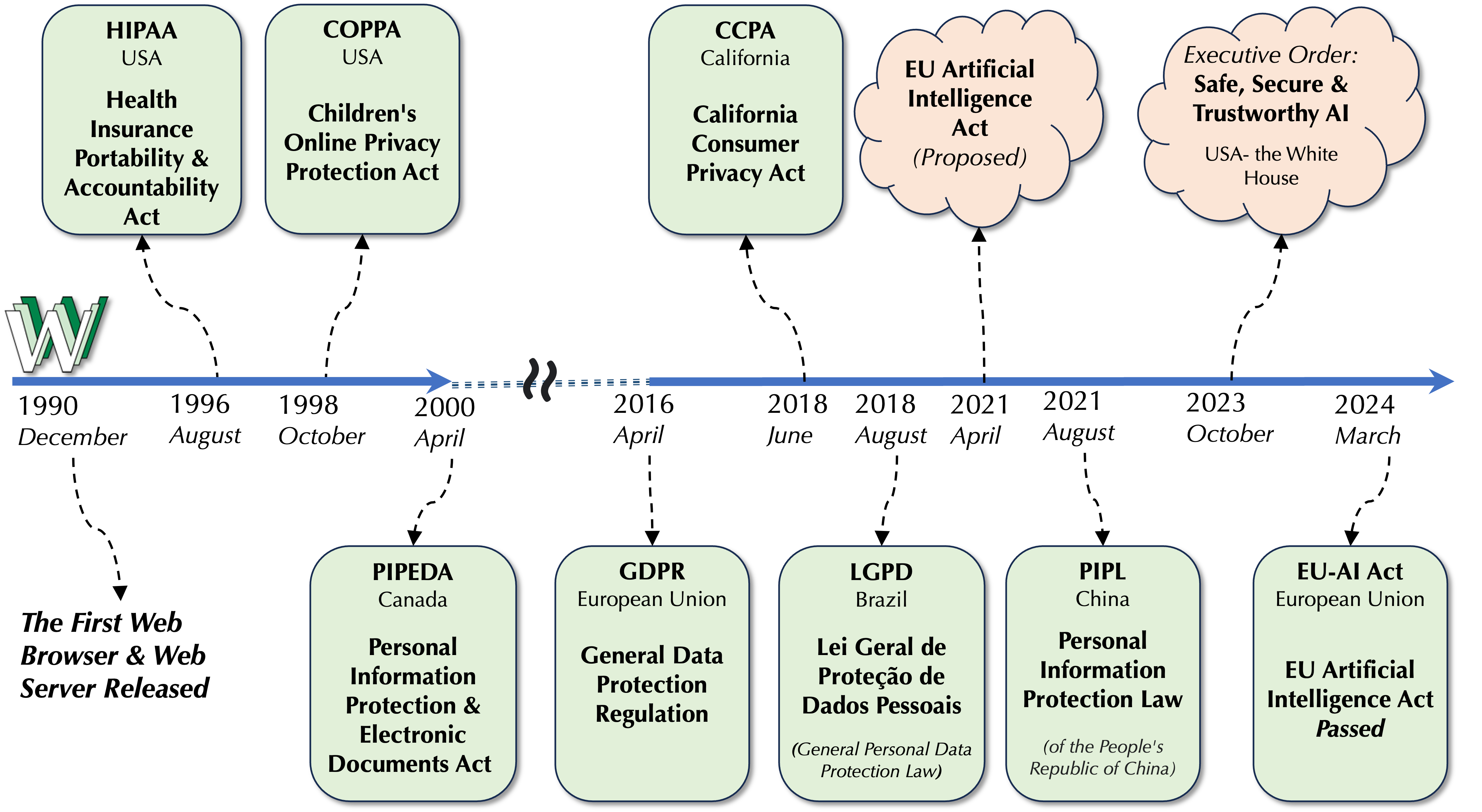}
    \caption{Timeline of data privacy laws passed by example legislative entities. While many countries have not yet passed or enacted their digital data privacy laws, pressured by public opinion, many legislative bodies will have completed their efforts to regulate AI within the next few years. For reference, in 1990, the first web sever and web browser were created by \textit{\href{https://en.wikipedia.org/w/index.php?title=Tim_Berners-Lee&oldid=1210713982}{Sir Tim Berners-Lee}}.}
    \label{fig:data-privacy-timeline}
\end{figure}

\begin{figure}
    \centering
    \includegraphics[width=0.85\linewidth]{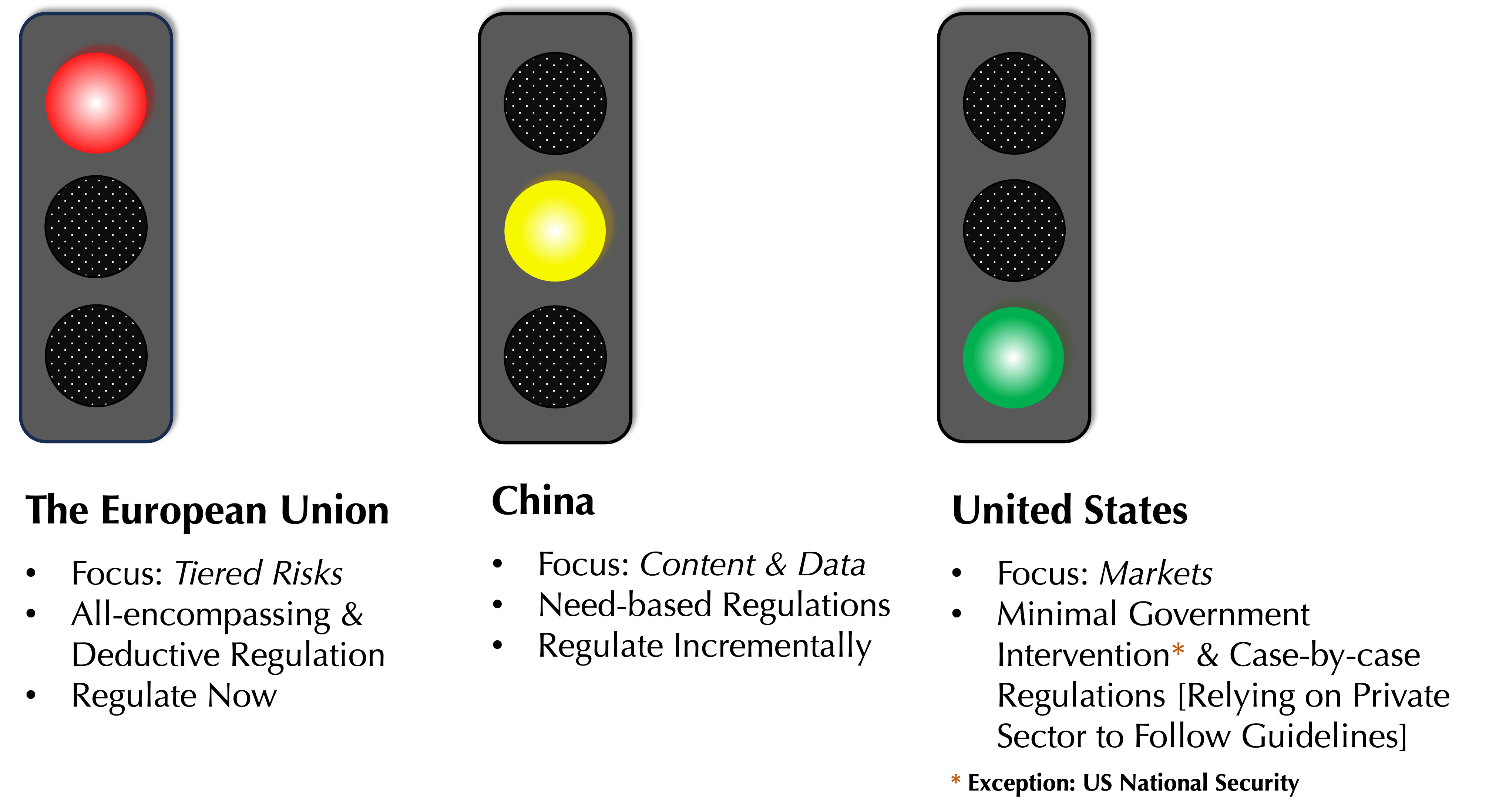}
    \caption{A high-level understanding of EU, USA, and China's legal viewpoint towards regulating AI. Here, \textbf{Stop}, \textbf{Caution}, or \textbf{Go} are various responses to the important question: `\textit{What should one do if existing laws may not have the capacity to regulate AI?}'.}
    \label{fig:us-eu-china-trafficlight}
\end{figure}

\noindent
In summary, varying approaches to AI governance at regional and international scales are shaped by factors such as political and technological leadership, data sovereignty laws, cybersecurity threats, cultural as well as ethical perspectives on AI use (see \cref{fig:us-eu-china-trafficlight}.

\subsection{AI Regulation Modes: Bottom-up \textit{vs} Top-down Development}
\label{ssec:bottom-up-vs-top-down}

As far as the modality of AI regulation is concerned, governments and international organizations have been experimenting with different implementation approaches. Common frameworks on regulation and governance are as follows:
\begin{enumerate}[label=\ding{43}]
    \item \textbf{Top-down Regulation:} Rules set by higher authorities or central government, trickle down to ensure compliance. It is widely used across sectors like finance, healthcare, and telecommunications to maintain order and public safety. Critics of this approach argue it stifles innovation and adaptability (\citet{homsy2019multilevel}). 
    \item \textbf{Bottom-up Regulation:} In contrast to `top-down regulation', this approach starts from local communities and governance. It heavily relies on self-regulation as well as community governance driven by grassroots organizations and independent entities. The `flow' of regulation is, therefore, upwards with higher authorities adopting the collectively verified policies (\citet{capano2012governance}).
    \item \textbf{Multi-level Regulation \& Governance:} Multilevel governance recognizes that policies and rules have to be flexible enough to be adopted at various levels, e.g. local, regional, national, and international levels. It involves coordination and cooperation among these different levels of government, as well as with non-state actors, to address issues that may cross traditional jurisdictional boundaries, e.g. mitigating risks of climate change (\citet{tortola2017clarifying}).
    \item \textbf{Other Forms:} For instance, \textit{Market-based Regulation}, \textit{Self-regulation}, \textit{Horizontal Regulation \& Governance}, \textit{Network-based Governance}, \textit{Democratic Governance}, and \textit{Hybrid Modes} are a few examples. For a review, we refer reader to \citet{levi2012oxford}).
\end{enumerate}

\noindent 
One major concern raised is how innovation in AI innovation may be impacted by the choice above? There is a clear \textbf{\textit{trade-off}} between the level of regulation\footnote{In general, the level of risk a government is willing to tolerate drive the strictness of regulation.} and innovation (\citet{chanbalancing}). In other words, a top-down approach may seem a natural way to start regulating AI where a central governmental entity `\textit{defines and controls}' enforcement. We have observed signs of such viewpoints in EU-AI Act requesting permission to use AI in certain `high-risk' domains (see \cref{sec:AI-regulation-by-countries}). Alternatively, a bottom-up governance heavily relies on private sector to ``\textit{self-regulate}'' and follow `best practices' in AI products and ecosystem\footnote{Critics of this approach point to recurring failures of ``big tech'' companies in self-regulation. For example, in 2017, Equifax data was hacked and sensitive data including credit history of more than 148 million Americans was stolen. Hackers exploited a known vulnerability in Equifax' software systems to access its database systems. It is reported that the security team at Equifax had failed to fix this issue despite having access to software patch two months prior to the incident (\citet{usatoday2017equifax})}.

\subsubsection{A few Open Questions}
\label{sssec:top-bottom-a-few-questions-to-decide}
For policy makers or communities aiming to be involved in regulation of AI and developing TAI frameworks, answers on the following ought to be considered:
\begin{enumerate}[label=\ding{43}]
    \item Should TAI and its legislation be based on a top-down, bottom-up, or market first?
    \item Can we prioritize \textit{bottom-up} strategy and involve STEM academics, social scientists, and legal scholars to lead debates and building the legal framework?
    \item Alternatively, prioritize government's role and authority in passing TAI regulations.
    \item Should any TAI framework be accepted and adopted within international communities, first?
    \item Should regulation of AI be approached through only a lens of `\textit{risk}', `\textit{security}', `\textit{national security}', `\textit{social/criminal justice}', `\textit{commerce}', `\textit{human rights}', `\textit{social prosperity}', `\textit{existential threat to humanity}', etc.? 
    \item Should regulators, scholars, consumers, and companies assume that in the not-so-distant future, AI products may exhibit \textit{agency} over their interactions with the digital and/or the physical world?
    \item In the near future, should having \textit{open access} to education as well as resources to build or use `AI-widgets' be considered a civil or a human right? \footnote{For example, in 2021, the United Nations passed a resolution titled \textit{``The promotion, protection and enjoyment of human rights on the Internet''}.(\citet{UNassembly2021InternetRight}).}
\end{enumerate}

\section{AI Regulation: Current Global Landscape}
\label{sec:AI-regulation-by-countries}

\begin{myDisclaimer}
\updateOnWHEO
\end{myDisclaimer}

The regulation of AI has been one of the top news topics of the last two years. So, how did this all start? We take one step back and first check the timeline of how several countries responded to \textit{online data privacy laws} after internet was born. As shown in \cref{fig:data-privacy-timeline}), many countries have recently passed or enacted online data privacy laws. We argue that a significant technical and legal debt owed to AI regulation is because of the challenges associated with enforcing digital data privacy laws that are either immature, ineffective, or nonexistent. 

In this section, we focus on the USA, China, and EU's recent announcements and legal activities to regulate AI. Due to the fast pace and rapid development of AI technology, no single country has concluded their AI regulation journey yet. Needless to say, it is easy to recognize different philosophical viewpoints to adopt and regulate AI, cf. \citet{capraro2023impact}. In the remainder, we provide more details on the latest efforts carried out by main players of AI and the potential implications for the private sector. 

\subsection{The United States of America: President Biden's Executive Order on `AI Safety'}
\label{ssec:president-biden-eo}
\begin{myDisclaimer}
\updateOnWHEO
\end{myDisclaimer}

On October 30th, 2023, the White House published an \href{https://www.federalregister.gov/d/2023-24283}{executive order}\footnote{\textbf{Update}: Unfortunately, the White House digital link to Executive Order 14110 has recently been removed. In addition to the US Government National Archive (\cite{BidenAIEO2023}), an online digital backup of this order can be accessed here: \cite{BidenAIEO2023UCSB}.} titled `\textit{Safe, Secure, and Trustworthy Development and Use of Artificial Intelligence}':

\begin{quote}
\textit{My Administration places the highest urgency on governing the \textbf{development and use of AI safely and responsibly}, and is therefore advancing a coordinated, Federal Government-wide approach to doing so.  The rapid speed at which AI capabilities are advancing compels the \textbf{United States to lead} in this moment for \textbf{the sake of our security, economy, and society}.} \cite{BidenAIEO2023}
\end{quote}

\noindent
This government-wide executive order explicitly directs over fifty government entities to devise 
and implement actions requested by the \newabbr{WHEO}{White House Executive Order}, a.k.a. EO 14110. 
\begin{figure}[htpb!]
    \centering
    \includegraphics[width=0.9\linewidth]{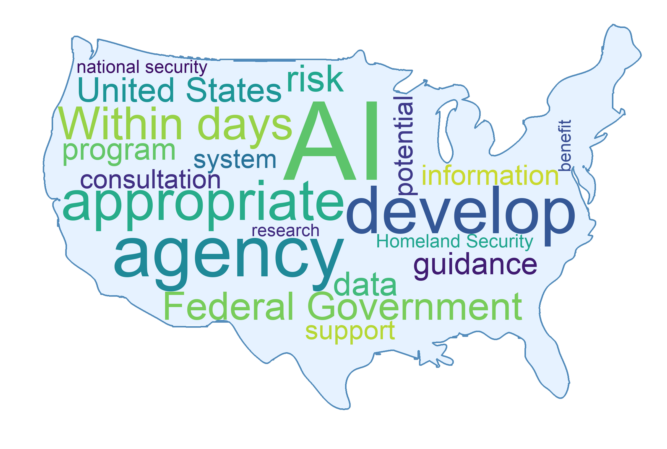}
    \caption{Word cloud shown above created using President Biden's executive order (WHEO-14110) 
    titled ``\textit{The Safe, Secure, and Trustworthy Development and Use of Artificial Intelligence}'' (\citet{BidenAIEO2023}). Larger words denote indicate that they are used more frequently in the text of  WHEO-14110.}
    \label{fig:wordcloud}
\end{figure}

\noindent
More specifically, WHEO-14110 aims to address \textbf{eight} overarching policy domains,

\begin{enumerate}[label=\ding{43}]
    \item Safety and Security,
    \item Innovation and Competition,
    \item Worker Support,
    \item AI Bias and Civil Rights,
    \item Consumer Protection,
    \item Privacy,
    \item Federal Government's use of AI,
    \item International Leadership,
\end{enumerate}

\noindent
by directing more than 50 US agencies to adopt and implement specific tasks as well as appropriate guidelines in a short period\footnote{Note that the phrase `\textbf{Within days}' is ranked as top 10 most frequently used term in the WHEO-14110. We have created a word cloud of the raw text and shown in \cref{fig:wordcloud}.}.

\subsection{The European Union: EU-AI Act}
\label{ssec:EU-regulations}

Commonly known as the `\textbf{EU-AI Act}', the European Union recently passed a comprehensive law\footnote{The full legal document of EU-AI Act can be found here: \citet{EUAIActRegulation2024}. A summary can also be accessed here: \citet{EUAIActSummary2024}.} to regulate AI products. To date, EU-AI Act is the only \textit{horizontal} legal framework towards regulating AI on such scale. At its core, development and deployment of application or services using AI must be categorized from a `\textit{risk management}' perspective (see figure \ref{fig:EUAIAct-RiskFlowChart}). The five risk categories defined by this law are as follows:

\begin{figure}[htp]
    \centering
    \includegraphics[width=0.99\linewidth]{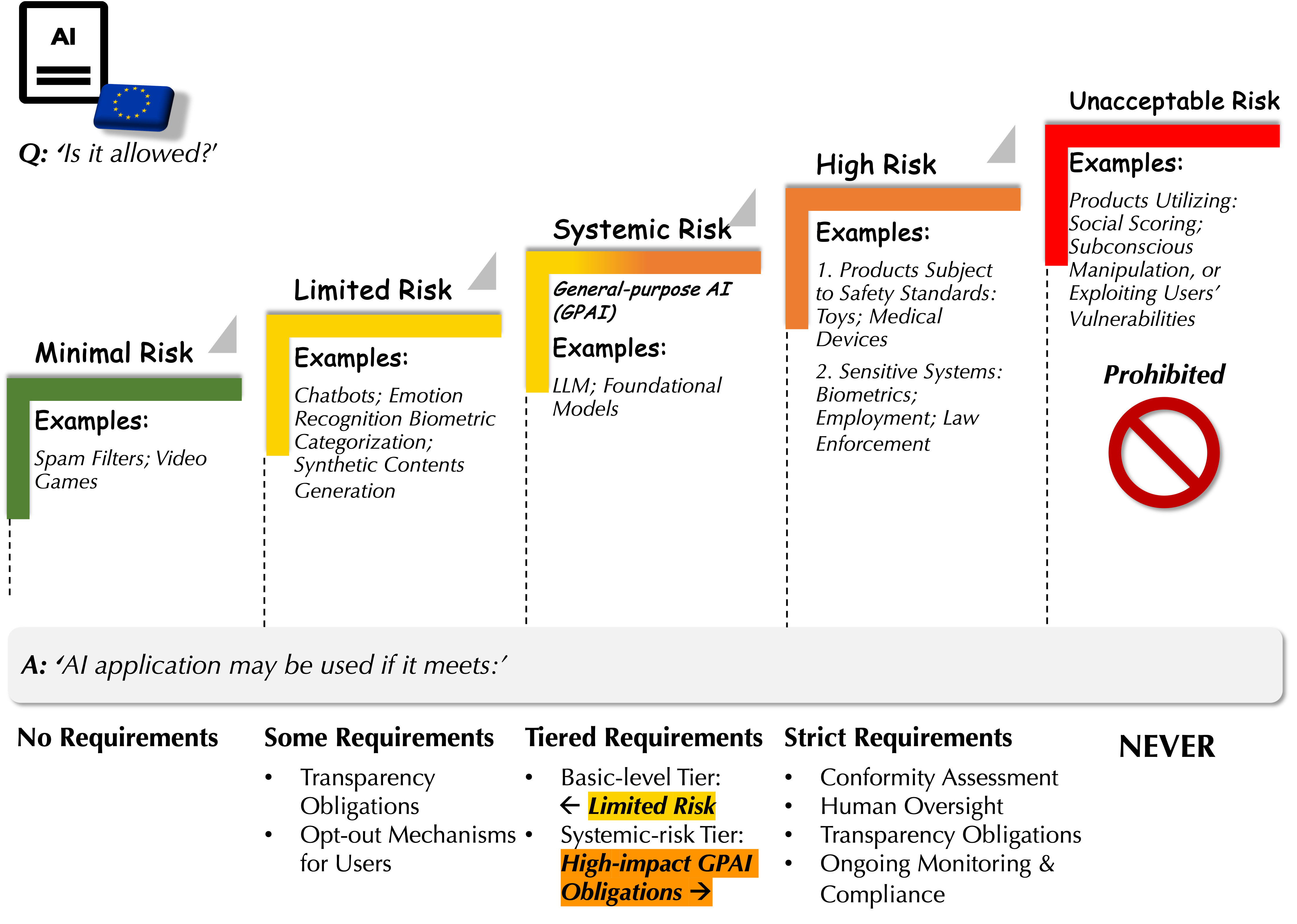}
    \caption{Schematics of EU-AI Act passed in March 2024. This risk-based regulatory approach states that any \mbox{AI-powered} application or service must first be categorized in one of the five predefined risk levels. For every risk level, any entity offering or using AI products must abide by requirements requested by the EU. For a summary of EU-AI Act, cf. \cite{EUAIActSummary2024}.}
    \label{fig:EUAIAct-RiskFlowChart}
\end{figure}

\noindent

\begin{enumerate}
    \item \textbf{Unacceptable Risk:} AI systems or products that are assumed to be hazardous to individuals and are banned. Common examples are social scoring systems, manipulative and subliminal products.
    \item \textbf{High Risk:} AI systems or products which could have negative impact on: a) safety; b) fundamental human rights
    \item \textbf{Limited Risk}: AI systems or products used to create or manipulate contents for human users (e.g. \href{https://en.wikipedia.org/wiki/Deepfake}{DeepFakes}, cf. \cite{westerlund2019emergence}), e.g. audio, video or image. 
    \item \textbf{Minimal Risk:} AI applications such as spam filters and video games are examples of AI applications which pose minimal risk to human user.
    \item \textbf{\newabbr{GPAI}{General-purpose Artificial Intelligence}:} AI products or systems that are built using `Foundational Models'\footnote{First popularized by Stanford Institute for Human-Centered Artificial Intelligence (HAI), a foundation (AI) model is a class of machine learning model (pre)trained to perform a range of tasks with minimal to null tuning effort.}, cf. \citet{zhou2023comprehensive,schneider2024foundation}. This risk category states that any AI product that is categorized as GPAI, inherently, has risk. Amendments further divide this risk into two sub-levels of risk demanding a set of additional requirements\footnote{It should be noted that this category was not in the original draft of EU AI Act and was added in a later version in 2023 due to rapid emergence of `Generative AI' products, e.g. ChatGPT or DALL-E.}.
\end{enumerate}

\noindent 
Penalties for violation depend on the company size and nature of violation, a) \euro35m (or up to 7\% of \newabbr{GAT}{Global Annual Turnover}) for prohibited violations, b) \euro15m (or up to 3\% of GAT) for majority of other violations, and c) \euro7.5m (or up to 1\% of GAT) for providing incorrect information to notified bodies (for more details, see \citet{euaiactFines}). 

\subsection{China}
\label{ssec:china-regulations}

China has undertaken a hybrid approach towards regulating AI. As China have focused on the most ``pressing'' and ``critical domains'' to be regulated first, i.e. social media, online contents, and recommendation engine. In doing so, as of today, three major regulations
\begin{enumerate}
    \item The Regulation of ``\textbf{Recommendation Algorithms}''. Issued on December 2021 (\cite{yang2022new})
    \item The Regulation of ``\textbf{Synthetic Content}''. Issued on November 2022 (\cite{sheehan2023china})
    \item Interim Measures for the Management of ``\textbf{Generative Artificial Intelligence}'' Services\footnote{By many, this law is considered a `\textit{breakthrough}' since it is the first international regulation pertaining to `Generative AI' technology.}. Issued on July 2023 (\cite{translate2023interim})
\end{enumerate}

\noindent 
have been implemented. While there is a rich lesson to be learned from China's path towards regulating AI, especially with respect to overcoming technical challenges associated with AI regulation, we remark that China's central government clearly mandates its politically-motivated requirements to be at the core of any AI regulation solutions. For example, `Article 4-- Requirement 1' regulation for Generative AI, one reads:
\begin{quote}
``\textit{...\textbf{Content generated through the use of generative AI shall reflect the Socialist Core Values}, and may not contain: subversion of state power; overturning of the socialist system; incitement of separatism; harm to national unity; propagation of terrorism or extremism; propagation of ethnic hatred or ethnic discrimination; violent, obscene, or sexual information; false information; as well as content that may upset economic order or social order.}''    
\end{quote}

\subsection{Other Countries}
\label{ssec:other-countries}

Almost every country had embarked on `AI regulation path' before the emergence of powerful \newabbr{GAI}{Generative Artificial Intelligence} systems. Prior to GAI, it would seem `reasonable' to assume there would be ample time for policy makers architect and pass relevant laws. That is not the reality in a post GAI world. Many leaders are now allocating public funding to research and development in TAI and AI safety. Balancing the trade-off between tight-grip regulation and innovation given political, economic, and sovereignty factors is an `\textit{art}'. 
\begin{enumerate}
    \item \textbf{UK:} United Kingdom's draft on AI regulation was first released in March 2023. UK's government clearly states a `pro-innovation' approach towards AI, \cite{govUKaiWhitePaper}. It states that unlike EU AI Act, UK government would not seek new government units and regulators for TAI. 
    \item \textbf{Japan:} has taken a `soft' approach towards TAI, i.e. no new regulation has been passed specifically to address TAI. Developers and companies should abide by existing and ``closest'' laws in data, software, and copyright. In a surprising move, Japan recently announced that use of copyrighted material to train AI models is permitted by law, cf. \cite{japancopyrightAI}.
    \item \textbf{Brazil:} Inspired by EU-AI Act, Brazil focuses on a risk-based approach towards regulating AI. In particular, it focuses on the rights of users interacting with AI systems from knowing that they are interacting with an AI agent, demand explanation, or even contest the decisions made by an AI system, especially for high-risk cases such as financial evaluations, cf. \cite{HolisticAIBrazil2024}.
\end{enumerate}

\subsection{What can be Learned from China, EU, and USA's Vastly Different Approaches to Regulate AI?}
\label{ssec:what-can-be-learned-from-china-eu-us}

\begin{enumerate}[label=\ding{43}]
    \item EU and China may face similar challenges in balancing trade-off between `control' and `innovation'
    \item China has taken the lead on drafting the first international regulation of \textbf{Generative AI}. 
    \item While not clear from day one, USA's current path towards AI regulation seems to support making AI openly and widely available, resulting in calls for more support of open-source platforms.
    \item EU's horizontal and deductive view towards AI regulation may seem restrictive. It has been criticized by several member states, e.g. France whose startup industry has been booming on AI and Generative AI. 
    \item One main benefit of EU's method is that it offers the benefit of longer-term planning and stability for private sector, as frequent updates to the EU-AI Act would not be necessary. Compare this to China's incremental legislation of TAI. 
    \item In contrast, as for USA and given the precedent-based justice and court system, it is a tedious task to ``anticipate'' the potential legal shifting landscape via local, state, or federal's perspective. 
\end{enumerate}

\subsection{How about Copyright?}
\label{ssec:copyright}

The significant success of recent AI models is owed to the abundance of publicly available and human-generated datasets. As an example \cref{tab:llama-training-dataset} provides the breakdown of different data sources and their respective sampling percentage used to train a language model named LLaMA-1\footnote{\newabbr{LLaMA}{Large Language Model Meta AI} is a foundational language model trained on 1.4 trillion tokens. It was first released on February 2023 by company Meta.}. In a `\textit{poetic}' way, these were the sources of education for LLaMA-1 to become ``\textit{literate}'' on human language.

Such digital datasets have become \textit{de facto} `source' of `public knowledge' representing human society at a global scale. Therefore, numerous none- and for-profit companies have been crawling, aggregating, and packaging them frequently. For instance, and as of April 5th 2024, there are close to 7 million articles on Wikipedia with over 4.5 billion English words (\citet{wikipediastats}). Another source of freely available content is \textit{arXiv.org} managed by Cornell University. ArXiv is a free distribution service that hosts nearly 2.4 million publications in scientific areas such as physics, mathematics, computer science, machine learning \& artificial intelligence, engineering, and economics.
\begin{table}[htbp]
  \centering
  \caption{Breakdown of dataset used to train a Gen-AI language model named LLaMA-1.}
  \label{tab:llama-training-dataset}
 \begin{threeparttable}
\begin{tabular}{l|cccc}
    \hline
    \hline 
    \rowcolor{gray!30}
    \textbf{Source Name} & \textbf{Topic} & \textbf{Sampling \%} & \textbf{Size on Disk} \\
    \hline
    CommonCrawl & Internet Websites \& Datasets & 67\% & 3.3 TB \\
    C4\tnote{*} & Clean English Text (Web) & 15\% & 783 GB \\
    GitHub & Programming & 4.5\% & 328 GB \\
    Wikipedia & General Knowledge & 4.5\% & 83 GB \\
    Books & History, Literature, Novels & 4.5\% & 85 GB \\
    ArXiv & Scientific Publications & 2.5\% & 92 GB \\
    StackExchange & Q\&A Websites on various Topics& 2\% & 78 GB \\
    \hline
  \end{tabular}

\begin{flushright}
\begin{tablenotes}\footnotesize
\item[*] `Colossal Clean Crawled Corpus' (C4) is clean English text extracted from the internet. For details, we refer the reader to \citet{raffel2020exploring}.
\end{tablenotes}
\end{flushright}
\end{threeparttable}
\end{table}
\noindent
\noindent
Automated web crawling tools ought to collect dataset from internet while honoring any copyright, privacy or other legal considerations. The reality is assuming no malicious (human) intent, such tools are never fully perfect. Their usage has invited conversations surrounding legal, ethical, safety/privacy, and copyright issues(cf., \citet{krotov2020tutorial}). 

Using such massive datasets to train generative AI models, with minimal to zero due diligence, has caused legal and PR incidents. For example, OpenAI is faced by two lawsuits (\citet{rawStoryVsopenai,interceptVsopenai}) filed by media companies on copyright violations. These cases are only the beginning of an anticipated barrage of lawsuits against companies that have trained or used generative AI models using such datasets. Ambiguity surrounding the topic of \textit{Copyright} and Generative AI can be exemplified as
\begin{enumerate}[label=\ding{43}]
    \item Can AI models be trained on copyrighted material?
    \item Can existing laws on `\textit{fair use of copyrighted material}' and `\textit{derivative work}' be applied to resolve legal disputes on AI training?
    \item Who owns the IP and the rights to outputs of an AI product? For example, in 2019, a Chinese district court recognized that the AI developer should hold the right to a news article created by an AI-enable robot. In contrast, in 2020 the patent and trademark offices in the UK, EU and the United States rejected patent applications where an AI-powered system was designated as an inventor (or co-inventor), cf. \citet{sun2021redesigning}.
\end{enumerate}

Diving deep into the complexities and nuances of generative AI and Copyright is beyond the scope of current paper. We conclude by stating that considering current legal and technical landscape, many scholars suggest a co-evolution of copyright laws along with proper data-governance technologies which can enable traceability, data quality, and `algorithmic unlearning' as an added requirement for any AI model, cf. \citet{yang2024generative,chu2024protect,gillotte2019copyright,henderson2023foundation,sag2023copyright,lucchi2023chatgpt,sobel2017artificial}.

\section{Risk}
\label{sec:risk}

In March 2022, the Arizona Supreme Court ruled that (\cite{arizonaCase}) the family of a 4-year-old girl named \textit{Vivian Varela}, who had been killed in 2015 in a car accident, can sue Fiat Chrysler Automobiles, the parent company of Jeep, for `\textbf{wrongful death}'. The family had argued that the automatic emergency braking system, which could have potentially prevented the crash, was not installed in the 2014-Jeep Grand Cherokee that rear-ended their car. Despite the availability of this life-saving technology\footnote{Multiple studies have reported 40\% to 70\% fewer rear-end and front end crashes, cf. \cite{cicchino2018real},  \cite{aukema2023real}, and \cite{fildes2015effectiveness}}, at the time, it was only offered as an optional feature bundled with a ``\textit{luxury package}'' for an additional \$10,000.

\noindent
In hindsight, the fatal crash could have been prevented if companies prioritized safety over profits. Jeep's decision to treat the `\textit{emergency braking system}' as a financial incentive rather than a standard safety feature reflects a misguided approach\footnote{In 2014, installing an emergency braking system was not a governmental mandate. Therefore, Jeep treated it as a `luxury' feature ought to be purchased by customers.}. 

AI-enabled decision-making tools, sometimes referred to as `digital twins', are becoming integral parts to various fields including engineering, business, human resources, procurement, and government. As their use continues to proliferate, we expect an escalation in the complexities surrounding ethics, engineering, and profitability. In extreme instances, the legal implications may thrust any court/justice system into uncharted territory, potentially, establishing new legal precedents.

\begin{figure}[htp]
    \centering
    \includegraphics[width=0.8\linewidth]{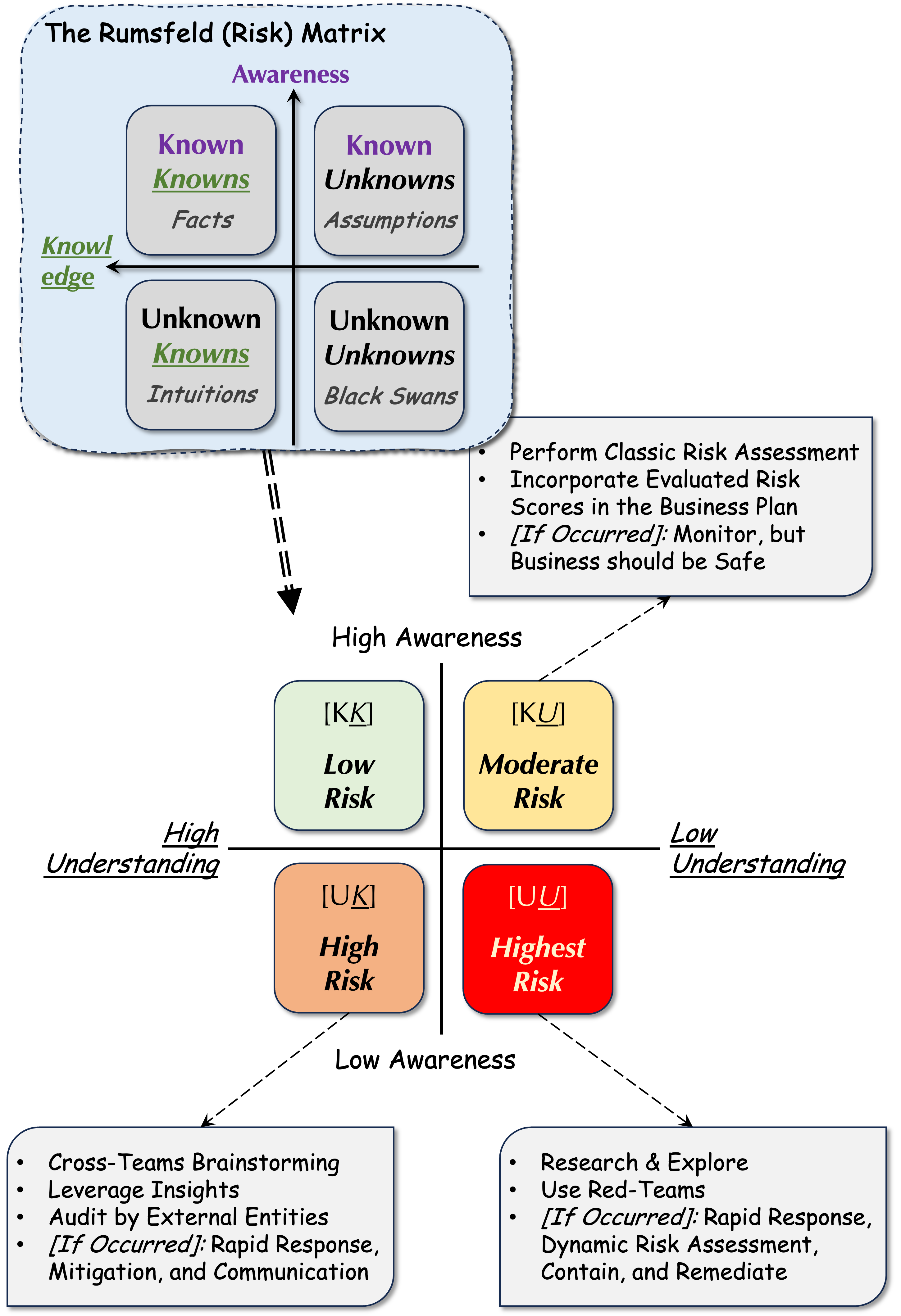}
    \caption{Risk quadrants (also known as the Rumsfeld Risk Matrix (RRM)) and common recommended action for each risk level. Here, U\textit{U}, U\textit{K}, K\textit{U}, and K\textit{K} refer to \UU, \UK, \KU, and \KK\quad respectively. It is important to consider the action plans to mitigate risk according to each region. RRM can be employed by any team building or using an AI system to plan for, mitigate, or remediate potential risks or legal challenges.}
    \label{fig:risk-quadrants}
\end{figure}

\noindent

\subsection{Managing Risk and Making \textit{Good} Decisions under Uncertainty}
\label{ssec:navig-risk-uncertainty}
Within any organization, managers and decision makers are expected to understand, plan, mitigate, and navigate risks. Disciplines such as `Operations Research', \newabbr{ERM}{Enterprise Risk Management} (cf. \citet{bromiley2015enterprise}), `Strategic Management', are only a few examples. 
In general, such disciplines aim to combine structured, empirical, and statistical frameworks so that managers facing uncertainty, could plan for risks or make informed decisions. Often, uncertainty is rooted in having an incomplete view/data into the status of company, product, demand, clients, customer behavior, or true randomness, also known as `\textbf{\textit{aleatory uncertainty}}'\footnote{Aleatory uncertainty refers to inherent randomness in a phenomenon that can never be predicted, e.g. outcome of a (fair) coin toss. In contrast, `\textit{epistemic uncertainty}' (also known as `\textit{systematic uncertainty}') refers to inaccuracies in data or observations that can be reduced or eliminated by means of more experiments or collecting new data. For a review on aleatory and epistemic uncertainty, cf. \citet{der2009aleatory}.}.

In the context of TAI, it is important to recognize how every category of uncertainty can be estimated, measured, detected, reduced, or eliminated. Transforming such `unknowns' into `risk score' or `risk level' compatible with existing ERM is a non-trivial task. While ERM for IT and Cybersecurity has been a well-studied discipline, to the best of our knowledge, there is no widely accepted framework to incorporate TAI in ERM for every organization.

As the first step, with the adoption of UQ by the AI scientific community, we now have access to several mathematical and statistical techniques estimating uncertainties associated with an AI model output (cf., \citet{hullermeier2021aleatoric,gawlikowski2021survey}).

As an example, consider a car with an `intelligent' \textit{automatic brake system}, which uses computer vision to detect nearby objects and preventing collisions. This module is, however, designed to operate to assist the human driver only in `normal or acceptable' conditions. When the driver is facing unfavorable conditions such as, extreme fog, this intelligent system \textbf{must} recognize its lack of `\textbf{\textit{confidence}}' in the outputs returned by the computer vision module, warn the driver, and disengage safely.

In this section, we select the \newabbr{RRM}{Rumsfeld Risk Matrix} and apply risk management in the context of TAI. We show how a simple framework such as RRM be incorporated into AI products or systems and map the risk categories associated with every step of an AI product life-cycle into `actionable' insight required for efficient implementation of TAI. 

\begin{figure}[htpb]
    \centering
\includegraphics[width=0.99\linewidth]{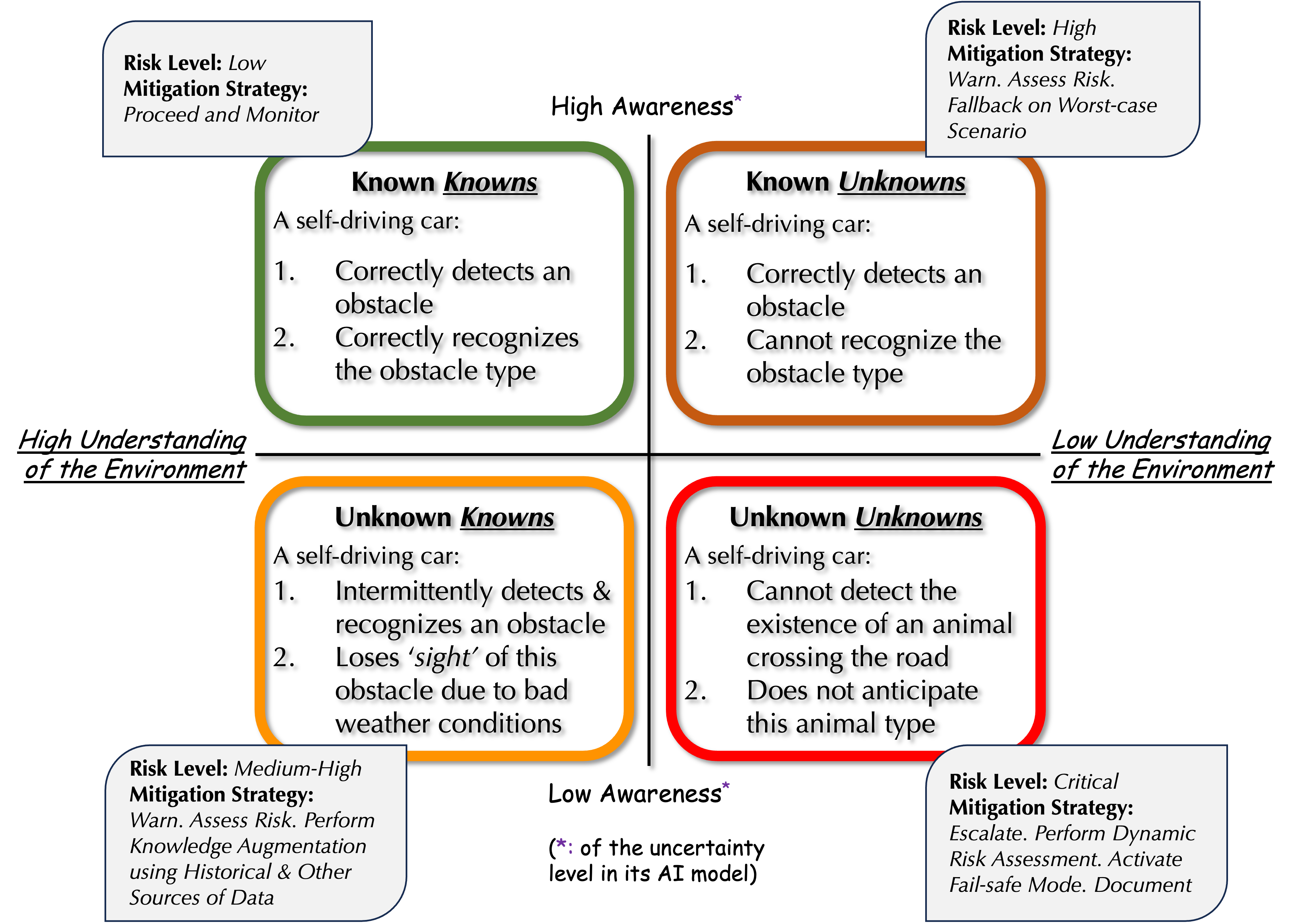}
    \caption{Rumsfeld Risk Matrix (RMM) constructed for a hypothetical scenario: An AI model utilized by a `self-driving car' to identify objects, humans, and animals on the road. This is an example of how using RMF during or after training and \textit{productionizing} an AI model system can benefit the engineering or test teams. Depending on the risk level, quadrant, managers, and stakeholders can estimate the risk associated with every quadrant, and adjust the AI model or mitigation resources, accordingly.}
    \label{fig:UKSelfDriving}
\end{figure}

\subsection{Example: Collecting Training Data and Mapping Risk to Actions}
\label{ssec:collcecting-training-data}

In the context of \textbf{collecting training data} to build a new AI product, the following are examples of risks and how they could fall under each quadrant in the RRM matrix. 

\begin{enumerate}[label=(\roman*)]
    \item \textbf{\KK:} Collecting user activity data-- training data-- from a \textit{biased} source, such as a social media platform where users are more likely to express extreme views.
        \begin{itemize}[label=\ding{43}]
            \item Source of training data is known to be unreliable or biased.
            \item Training data is inaccurate or incomplete.
            \item Training data includes sensitive information that could discriminate against certain groups of people.
        \end{itemize}
        \colorbox{lightgray}{\textit{Mitigation strategy:}} Diversify data sources to reduce the risk of bias.
    \item \textbf{\KU:} Collecting training data that is not perfectly accurate yet suffices for building an AI model.
        \begin{itemize}[label=\ding{43}]
            \item Collecting training data from a new or emerging source that has not been previously evaluated for quality or bias.
            \item Collecting training data from a source that is known to be reliable, but the specific data being collected has not been evaluated for quality or bias.
            \item Collecting training data that is known to be accurate and complete, but the potential impact of using the data to train an ML model is unknown.
        \end{itemize}
        \colorbox{lightgray}{\textit{Mitigation strategy:}} Apply data quality control measures to identify and correct-- if possible-- errors, e.g. missing data, in the data.
    \item \textbf{\UK:} Collecting training data that is highly sensitive and if not handled by experienced data scientists, could result in discrimination against certain groups of people.
        \begin{itemize}[label=\ding{43}]
            \item Collecting training data from a source that is unknown, but the likelihood of the data being biased or inaccurate is known to be high.
            \item Collecting training data from a source that is unknown, but the potential impact of using the data to train an ML model is known to be high.
        \end{itemize}
        \colorbox{lightgray}{\textit{Mitigation strategy:}} Implement strong data security measures to protect the data. This category of data can be only used based on per case-by-case and approval by \newabbr{CIO}{Chief Information Officer}
    \item \textbf{\UU:} Collecting training data which lacks proper \textit{meta-data}\footnote{Meta-data refers to a set of attributes which describes data at hand. For example, a photo taken by a smartphone may include meta-data such as GPS coordinates and time/date when taken.} or documentation to its original source. Therefore, it is not clear if this dataset is particularly relevant to the task which the new AI model will be used for.
        \begin{itemize}[label=\ding{43}]
            \item Collecting training data from a source that is completely unknown, and both the likelihood of the data being biased or inaccurate and the potential impact of using the data to train an ML model are unknown.
        \end{itemize}
        \colorbox{lightgray}{\textit{Mitigation strategy:}} Investigate further with other teams to find out the source of data. If applicable, conduct exploratory analysis in a protected and `sandbox' environment.
\end{enumerate}

\subsubsection{Web-crawled Datasets and their Unknown Risks}

In \cref{ssec:copyright}, we provided examples of web-crawled datasets (originated from internet) to train or fine-tune recent generative AI models. In this section, we discuss the potential risks associated with using such datasets with an emphasis on risk management. 

Collecting and aggregating dataset from internet automatically can pose direct and secondary risks. The imperfections in the dataset or cleansing tools would allow unwanted (or illegal) content be retained and later be used to train an AI model. The sheer size of such datasets (for example, see \cref{tab:llama-training-dataset}) renders any manual inspection, infeasible. Once a generative AI model is trained (on this imperfect dataset), secondary effects such as sharing sensitive data, producing Copyrighted material, or echoing inaccurate responses to a human user (aka \textit{hallucination}) can take place. These examples can pose risks in \UK\quad or \KU\quad quadrants (see RMM in \cref{fig:risk-quadrants}). Under these circumstances, and as part of a company's ERM framework, negative consequences can be either avoided (e.g. by purchasing sanitized dataset from trusted vendor), patched, unlearned, or be minimized via monitoring or red-teaming tests. 

Unfortunately, there are scenarios where using web-crawled dataset can cause unanticipated consequences of the nature that belongs to \UU-- the hardest to plan for and most expensive quadrant in RMM. As an example, in a recent study published by Stanford Internet Observatory (\citet{thiel2023csam}), publicly available image datasets crawled from internet (by a reputable non-profit company based in Germany \newabbr{LAION}{Large-scale Artificial Intelligence Open Network}), \textbf{LAION-5B} and \textbf{LAION-400M}, included verifiable links to \newabbr{CSAM}{Child Sexual Abuse Material}. Before release of this diagnostics, these datasets had been used to train popular \textit{text-to-image} AI models such as \textit{Stable Diffusion} (\citet{rombach2022high}) and Google's \textit{Imagen} (\citet{saharia2022photorealistic}). Currently, there are multiple efforts and call-to-actions to build safe frameworks in order collect, prepare, and validate the legal rights of AI datasets and their development cycle, cf. \citet{khan2022subjects,longpre2023data}.

\subsection{AI Regulatory Sandbox: A Useful and Interim Medium}
\label{ssec:regulatory-sandbox}

We firmly believe that we should use all the means to allow innovation in the AI domain alive. Provisions mentioned in the EU-AI Act and WHEO-- with reasonable intentions-- could ultimately stifle innovation as well as engagement at the community levels. We have yet to observe the actual implementation and guidelines-- as they say, the devil is in the details\footnote{If such provisions are not implemented tactfully, we believe it may lead to a state where only a few wealthy and resourceful conglomerates can ``afford'' the risks and subsequent legal fines provisioned in the EU-AI Act. In other words, individuals and startups driving any meaningful innovation in TAI are heavily discouraged.}

To mitigate this, EU-AI act introduces a new concept called `\textbf{AI Regulatory Sandbox}' which encourages the EU members to create regulatory environments, tools, and best practices for testing and experimentation with new AI products-- under supervision of EU members and approved authorities, cf. \cite{truby2022sandbox}. In essence AI regulatory sandbox serves two purposes:
\begin{enumerate}[label=\Roman*.]
    \item Foster learning and innovation in AI for businesses via real-world development and testing of new AI-powered products.
    \item Contribute to regulatory learning by creation and testing experimental legal frameworks around new technologies based on AI.
\end{enumerate}

\noindent
While this provision in EU-AI act has yet to be finalized, Spain has recently launched the first program of this kind to foster AI innovation while evaluating regulatory requirements to be enacted in EU-AI Act.  This point of view seems to be gaining a widespread interest as it aims to expand beyond EU. For instance, Sam Altman-- CEO of OpenAI-- recently invited the \newabbr{UAE}{United Arab Emirates} to become a testing ground for AI regulation (\citet{bloombergOpenai2024}). 



\section{Bias and Fairness}
\label{sec:fair-ai}

\begin{keytakeaways}
    \keypoint{There are multiple definitions for `fairness'.}
    \keypoint{Mathematically speaking, it is proved that not all aspects of fairness-- characterized by definitions-- be enforced concurrently.} 
\end{keytakeaways}

\subsection{`Biased AI': A Polysemic Term Which Needs Clarification}
\label{ssec:biased-ai-polysemic-term}

For better or worse, diverse community surrounding AI have been using the term `\textbf{biased}' AI to often disparate technical or conceptual topics. This may have caused unnecessary ambiguity and sometimes confusion, cf. \citet{felzmann2019transparency}. In order to `decode' this term, it is important to pause and reflect on two main clarifying items with respect to any biased AI system: 1) What is the context that AI product is used? and 2) Who is the SME and his/her role in AI life-cycle?. For instance, consider the following SMEs studying or mitigating bias in AI-enabled system:
\begin{enumerate}
    \item \textbf{AI Engineer/Data Scientist:} Algorithmic bias – The systematic error introduced by the design and implementation of machine learning algorithms. While an entirely mathematical concept, if not detected properly, it may result in unreliable or even unfair outcomes, cf. \citet{barocas2016big,kordzadeh2022algorithmic,belkin2019reconciling,curth2024u}.

    \item \textbf{Regulator/Policy Maker:} Social or human bias – The unfair and prejudicial treatment of certain individuals or minority groups usually caused by pre-existing societal and historical biases reflected in the data used to train AI models, cf. \citet{buolamwini2018gender,noseworthy2020assessing}.
    
    \item \textbf{Ethicist/Philosopher}: Ethical bias – The moral implications of AI decision-making, which may involve value judgments, unequal treatment, or perpetuating existing social inequalities, cf. \citet{hagendorff2022blind,jobin2019global,mittelstadt2016ethics}.
    
    \item \textbf{Data Analyst:} Statistical bias – The difference between an algorithm's expected prediction and the true value, which can result from errors in data collection, sampling, or modeling assumptions, cf. \citet{hastie2009elements}.
    
    \item \textbf{User Experience (UX) Designer:} Interaction bias – The biases that emerge from the design of AI interfaces and how users interact with them, potentially leading to unintended consequences or unequal access to AI-driven services, cf. \citet{yoon2023ethical,bach2022systematic,meske2020transparency,oh2018lead}.
    
    \item \textbf{Social Scientist:} Systemic bias – The ways in which AI systems can perpetuate and amplify broader social, economic, and political inequalities, cf. \citet{beer2019social,fountain2022moon}

    \item \textbf{Legal Scholar:} Legal bias – The potential for AI systems to generate outcomes that violate existing laws, regulations, or legal principles, such as those related to non-discrimination, privacy, or even due process\footnote{Legal scholars would be particularly interested in understanding how AI systems can be designed, implemented, and governed to ensure compliance with existing law and protect individuals' rights. They could also consider the challenges of holding AI systems and their creators accountable for biased outcomes, auditing AI systems without violating intellectual property rights, and the potential need for new legal frameworks to address these issues.}, cf. \citet{citron2014scored}.
\end{enumerate}
These interpretations demonstrate the \textit{polysemic} nature of the term 'Bias' in AI and ML, as its meaning can vary significantly depending on the context and the persona using it.

\subsection{Bias as State-of-mind of an Individual}
\label{ssec:human-bias}
\topquote[Decimus J. Juvenalis (circa 1st/2nd century CE); A Roman poet]{Quis custodiet ipsos custodes? (Who will watch the watchmen?)}

Human oversight is paramount in the end-to-end life-cycle of AI models. This also includes auditing and monitoring systems designed to deliver a medium for TAI. However, with humans within any organization, we anticipate that their decision-making process is not immune to multiple forms of cognitive bias in the human brain. Scholars in psychology have conducted an exhaustive search and indicated how this bias category could overshadow honest interpretation of any situation. More recently, there have been studies to harvest this knowledge from the field of psychology to tackle various forms of biases hurting AI systems, cf. \citet{tambe2019artificial,ashmore2021assuring}. Without going into details, below is a few common categories of biases known to impair human judgement:
\begin{enumerate}[label=\ding{43}]
    \item Cognitive Bias
    \item Confirmation Bias
    \item Anchoring Bias
    \item Ethical Fading
    \item Primacy Effect
    \item Group-think and Conformity Paradox
    \item Self-serving Bias
    \item Moral Licensing
\end{enumerate}

\noindent 
We note that understanding how the list above can impact SMEs in charge of overseeing or investigation potential problems within data, AI model, audits, testing, or quality assurance is a must. 
considering above list in drawing conclusion is key. For every bias type, and in the context of TAI, there are different mitigation strategies that can help decision makers and SMEs minimize the risk imposed by cognitive bias in handling of the TAI system. 

For example, in their investigation, \citet{de2019predicting} show that financial risk assessment reports conducted by various independent entities in Europe seemed to be depended on:
\begin{enumerate}
    \item The magnitude and impact of the financial catastrophe or scenario under review. 
    \item The total number of signatories of the produced report, i.e. only one individual (\emph{vs} more than one person signed the final report.
\end{enumerate} 

\noindent
\Citet{de2019predicting} conclude that the auditing firms seemed to be more concerned about the \textit{public reaction} to the company's reputation and, therefore, tried to avoid any potential media scandals because of their findings shared in public reports. While it may seem dire, in the context of TAI, there have been systematic approaches that can minimize or eliminate biased decision of the `\textit{watchmen}'. It is beyond the scope of the current paper to dive into such mitigation strategies. For more insight on this topic, we refer reader to \citet{wall2017warning,mohanani2018cognitive,kliegr2021review} and references therein.

\subsection{Fairness}
\label{ssec:fairness}

There seems to be rather universally accepted `conventional wisdom' which states that any \textit{mathematical} definition of `\textit{fairness}' is often at odds with \textit{human perception} of being (un)fairly treated at an individual level. In other words, a person can employ `\textit{relativistic}' or `\textit{contrastive}' arguments to justify why she `\textit{felt}' being treated unfairly (\citet{srivastava2019mathematical}). 

For example, consider this hypothetical situation; A passenger at an airport is asking to be upgraded to `business class'. After checking with the gate agent, her request is turned down. In her complaint later submitted to the airline customer service, this passenger states the following as grounds for being \textit{unfairly treated}:
\begin{quote}
    ``\textit{I have been a loyal customer of \textit{Almost-Landing Airlines}\footnote{We trust that our reader infers such names are completely fake and do not represent any particular airline.} for 10 years with a frequent-flyer status. \textbf{I was denied an upgrade} to \textit{business class}, even though there were plenty of empty seats. Also, I witnessed \textbf{another passenger without a frequent-flyer status being upgraded} with no issues. That is \textbf{not fair}...}''
\end{quote}

\noindent 
This example shows an individual perception of unfair treatment which may never exist 50 years from today. As human society advances, so do new definitions-- read perception-- of `fairness' (for an overview of how fairness metrics evolved in the past 50 years, cf. \citet{hutchinson201950}).

In order to quantify `algorithmic fairness', clear and commutable metrics denoting `fairness' must first be defined\footnote{More recent research also calls for approaching fairness in AI from a model perspective (as opposed to gauging \textit{fairness score} as an intrinsic property of an AI system), cf. \cite{lalor2024should}.}. Any mathematical definition of fairness must focus on measurements (data), clearly defined equations accompanied by the proper statistical framework, and various implications of measured fairness metrics in human-interpretable fashion.

\subsection{Widely Accepted Definitions for Fairness}

There is no unique metric which defines fairness. Any definition of fairness which is accepted by society\footnote{We remark that the notion of fairness not only depends on the existing cultural context, but it can also vary over long periods of time, cf. \cite{saxena2019fairness}. Consider how recently women were allowed to vote, even in first-world countries such as the US or Switzerland}, has yet to be transformed into mathematical or statistical manifestation. Once this is agreed up, any AI-system (or a statistical inference module) can be subjected to a `fairness assessment' compute engine which would quantify a `\textit{fairness score}'. This would be the first critical step to remediate any unintentional and unacceptable unfair decisions made by an AI-system. Alternatively, one can directly `\textit{map}' computed fairness score to a risk score (e.g. for the case of EU-AI-Act) and plan accordingly. Below is a widely accepted list of fairness metrics used by scholars in TAI research (collected from \cite{mehrabi2021survey} and references therein):
\begin{itemize}
    \item Equalized Odds (Group)
    \item Equal Opportunity (Group)
    \item Demographic Parity (Group)
    \item Fairness Through Awareness (Group)
    \item Fairness Through Unawareness (Group)
    \item Treatment Equality (Group)
    \item Test Fairness (Group)
    \item Subgroup Fairness (Subgroup)
    \item Counterfactual Fairness (Individual)
    \item Fairness in Relational Domains (Individual)
    \item Conditional Statistical Parity (Individual)
\end{itemize}

\subsection{Fairness Through the Lens of Group Size}
According to the number of individuals impacted by the outcome of an AI model, here are three classes of AI fairness enforcement\footnote{There are several reasons causing AI fairness \textit{violations}, e.g. training data size and quality, algorithms, AI model structure, poor AI model training and cross validation, data leakage, ignoring confounding features, are only a few. We invite our reader to recent survey \cite{caton2024fairness} and references therein.}:
\begin{enumerate}[label=\ding{43}]
    \item \textbf{Individual Fairness:} Aims for similar predictions (produced by AI model) for `\textit{similar}' individuals. For example, consider a male professor and a female professor with very similar financial history applying for a new `credit card'; they should both receive a similar spending credit line, e.g. \$10,000. In a \textit{fair system}, gender is considered a protected attribute and, therefore, should not determine solely the \textit{creditworthiness} of an applicant.
    
    \item \textbf{Group Fairness:} Consider company \textit{RandomBigCompany} launches a newly trained AI model to ingest candidate Resumes and suggest potential candidates for interview. Over a year, AI model reviewed more than 12,000 applications and suggested 500 cases for interview. Based on an internal audit, hiring manager noticed that the odds of male candidates being selected (by the AI model) over female candidates were consistently higher (3x), despite having similar qualifications. This is an example of group fairness violation.\footnote{While this is a hypothetical scenario with fake statistics, there have been similar problems reported, cf. \cite{amazon_bias_article}.} Considering the principle of \textit{group fairness}, any predictive model ought to be be adjusted so that applicants from groups\footnote{A group may be defined differently considering legal or policy requirements. For instance, according to legal mandates outlined in Title VII of Civil Rights Act of 1964 (\citet{act1964civil}), any AI model ought to be employed in hiring process should be designed such that a candidate's race, color, religion, sex, or national origin would not play a decisive factor in the outcome of employment.}, are given equal chances of being selected for interviews, assuming all other factors are equal.
    
    \item \textbf{Subgroup Fairness:} Combines both of the above viewpoints to investigate fairness. For example, in a model predicting recovery period for patients undergoing surgery, the parent group could be all patients over 65 years old. A subgroup within this group can be defined using indicators such as geographical location of patients, e.g. rural areas \emph{vs} urban areas. If the AI model disproportionately disadvantages\footnote{AI model may \textit{blindly} predict a lower likelihood of recovery for patients belonging to the rural areas, not because of their health status, but because they may have less access to healthcare resources, e.g. specialists or medical facilities.} patients from rural areas, it can signal a fairness violation at subgroup level. For an in-depth discussion on this topic, see \cite{kearns2018preventing}.
\end{enumerate}

\noindent 
To make matters even more challenging, there has been mathematical proof indicating that enforcing all definitions of fairness simultaneously is not possible, cf. \cite{kleinberg2016inherent}. 
\begin{figure}[htp]
    \centering
    \includegraphics[width=0.65\linewidth]{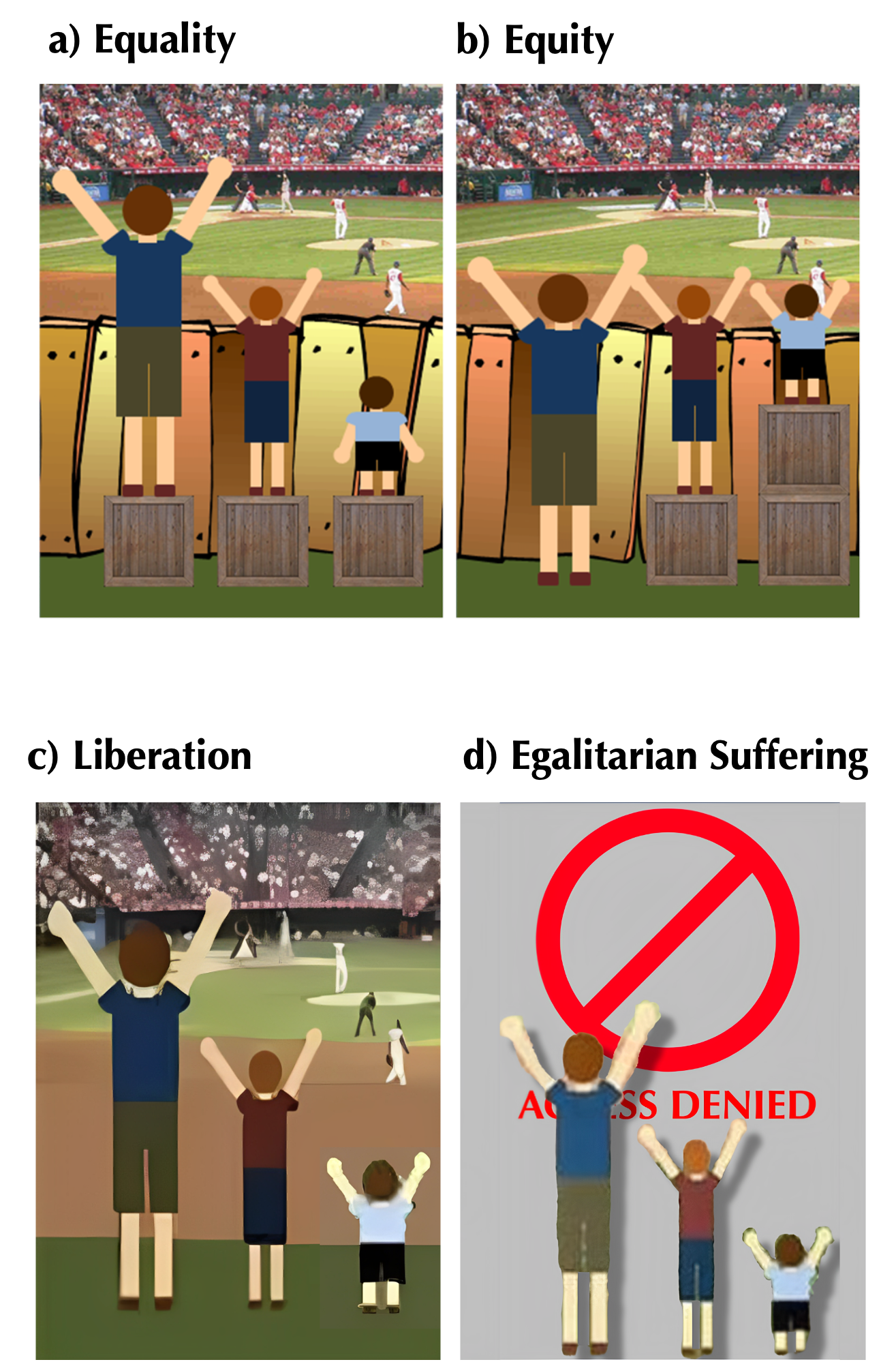}
    \caption{A simplified depiction of different definitions of `fairness' and associated interventions aimed to implement `justice'. (For a full story on the history of this meme, see \cite{equalityMeme2023})}
    \label{fig:equalityMeme}
\end{figure}

\subsection{AI Fairness and Human Rights: COMPAS Example}
\label{ssec:fairness-compas-example}

\newabbr{COMPAS}{Correctional Offender Management Profiling for Alternative Sanctions} built by a private company called \href{https://www.equivant.com/}{Equivant} has been used by U.S. courts to assess the likelihood of a defendant becoming a `\textit{recidivist}' in the next two years. Judges in the past have used this product to make decisions on defendant requests submitted to courts. While this topic has been widely debated amongst independent investigators, it is imperative to note how having limited to no-access to inner-workings, training data, as well as the audit results of software like COMPAS have eroded public trust. Below is a list of concerns raised by independent scholars: 
\begin{enumerate}[label=\ding{43}]
    \item \textbf{Human Rights:} COMPAS is a propriety product built by a for-profit company. Currently, it is closed to public scrutiny. Not only this violates 14th Amendment right, but COMPAS' lack of transparency has caused conflicting conclusions carried out by independent investigators, cf. \citet{rudin2020age}. 
    \item \textbf{Human \emph{vs} COMPAS:} In a recent study, it is shown that COMPAS does not perform better than an average conclusion derived from a pool of random strangers who are also not familiar with the criminal court systems, \cite{dressel2018accuracy}\footnote{In their study, authors recruited 400 volunteers. Every person was then asked to guess whether a \textit{defendant} would commit a crime within two years after studying a summary on defendants published by \href{https://www.propublica.org/}{ProPublica}.}. 
    \item \textbf{Racial Bias:}  COMPAS has caused controversy as it may violate `\href{https://www.archives.gov/milestone-documents/14th-amendment#:~:text=No%20State%20shall%20make%20or,equal%20protection%20of%20the%20laws.}{14th Amendment Equal Protection}' rights on the basis of race\footnote{Significant disparities in the recommendations returned by COMPAS software have been reported for \textit{African-American} and \textit{Caucasian} defendants (\citet{dressel2018accuracy}).}, since the algorithms are argued to be racially discriminatory with disparate treatment of African-American defendants, cf. \cite{thomas2022automating}. 
\end{enumerate}

\subsection{Our Proposed Solution: Example Template for `Fairness Verification and Validation Testing'}
\label{ssec:fairness-enforcement}

Let's assume a business has an overarching team, \newabbr{FVVT}{Fairness Verification and Validation Testing} who is responsible for the enforcement of fairness policy by product teams using AI models. 
FVVT team proceeds to engage and collaborate with several SMEs in order to:
\begin{enumerate}[label=\ding{43}]
    \item \textbf{Comprehending Legal Requirements:} Understand and manifest legal mandates and associated risks for business, customers, or other entities.
    \item \textbf{Identifying Impacted Product Teams:} If not clear, consider including every engineering team involved in data collection, training, deployment, and monitoring of an AI model.    
    \item \textbf{Operationalizing Legal Requirements}: a) Define or choose existing (Mathematical/Statistical) Fairness Metrics and KPIs. b) Map legal requirements into `acceptance criteria'. 
    \item \textbf{Select/Build Benchmarks:} Identifying relevant \textit{test} scenarios considering designated fairness dataset, AI model, and compute resources.
    \item \textbf{Test and Report:} Run agreed upon FVVT-benchmarks, report test results and metrics used, extract insights and report to various teams involved. If any violations observed, understand if current enterprise risk management strategy is capable of remediating such violations. 
    \item \textbf{Governance:} Versioning and document tests, testing dataset, selected benchmarks, human-readable interpretation of test results, AI models meta data, risk level, fairness KPIs.
    \end{enumerate}

\section{Explainable AI as an Enabler of Trustworthy AI}
\label{sec:xai}
\begin{figure}
    \centering
    \includegraphics[width=0.75\linewidth]{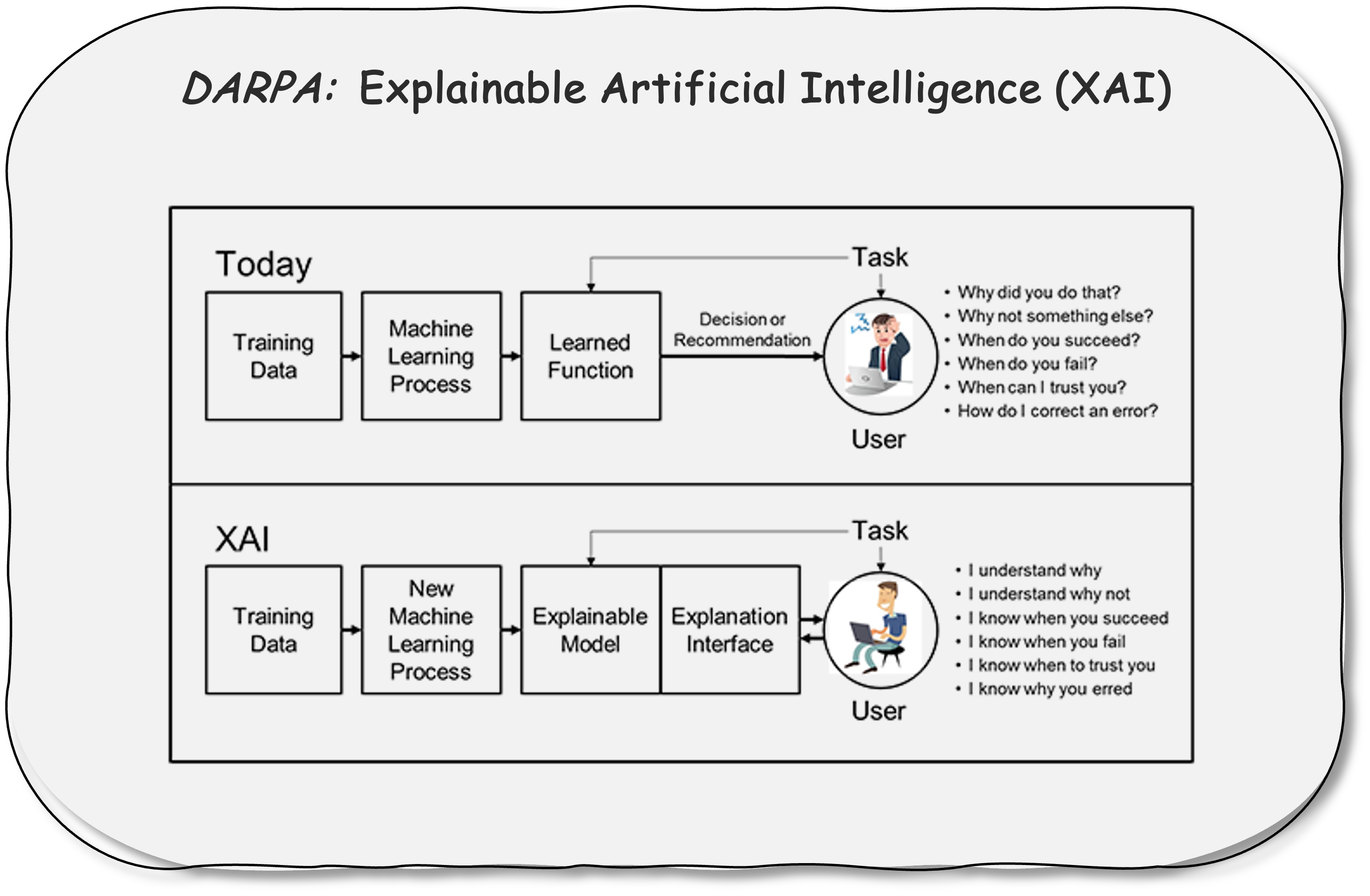}
    \caption{Schematics of XAI first introduced by DARPA in 2017, \citet{gunning2017explainable}.}
    \label{fig:xai-darpa}
\end{figure}

First coined by David Gunning\footnote{David Gunning, who is currently retired, was a program manager at \newabbr{DARPA}{Defense Advanced Research Projects Agency}.} in 2017 (see \cref{fig:xai-darpa}), the term `Explainable AI' has become a priority for research labs and companies. In DARPA's initial framework, XAI was to build machine learning and mathematical techniques which can be used to ``comprehend'' outputs produced by any black-box AI models, e.g. DNN. By doing so, an end user could rationalize `outputs' produced by an AI model before making (business) decisions. 

\noindent 
Over a short period of time, however, the expectations for XAI have significantly expanded in terms of diversity of end users as well as the level of details in XAI reports. With the expansion in scope, it is not surprising to see XAI research or taxonomy overlapping with other features of TAI. 

\subsection{XAI: Spectrum of Explainability and Interpretability}
\label{ssec:xai-specturum}

More adoption and `infusion' of AI-systems within Enterprise business cycles impose higher stakes for different end users. We argue that XAI tools and framework focused mostly on the realization of XAI for AI/ML experts. However, with the expansion of XAI to other stakeholders, stricter government regulations, and significant growth of black-box AI models, XAI is now different. 

Any organization implementing XAI in their business should expect that XAI ought to produce reports should consider main inputs: \textbf{WHO}, \textbf{WHEN}, and \textbf{WHY}. In other words, `explainability' component in XAI is dependent on the answers above. Hence, various stakeholders in an organization should expect a `\textit{spectrum of explainability}' to be considered. This spectrum has different levels of details, e.g. statistical terms to plain text, granularity of metrics across each attribute of TAI, or stratified level of access to the data or organizational chart. 

\begin{figure}[htp]
    \centering
    \includegraphics[width=0.99\linewidth]{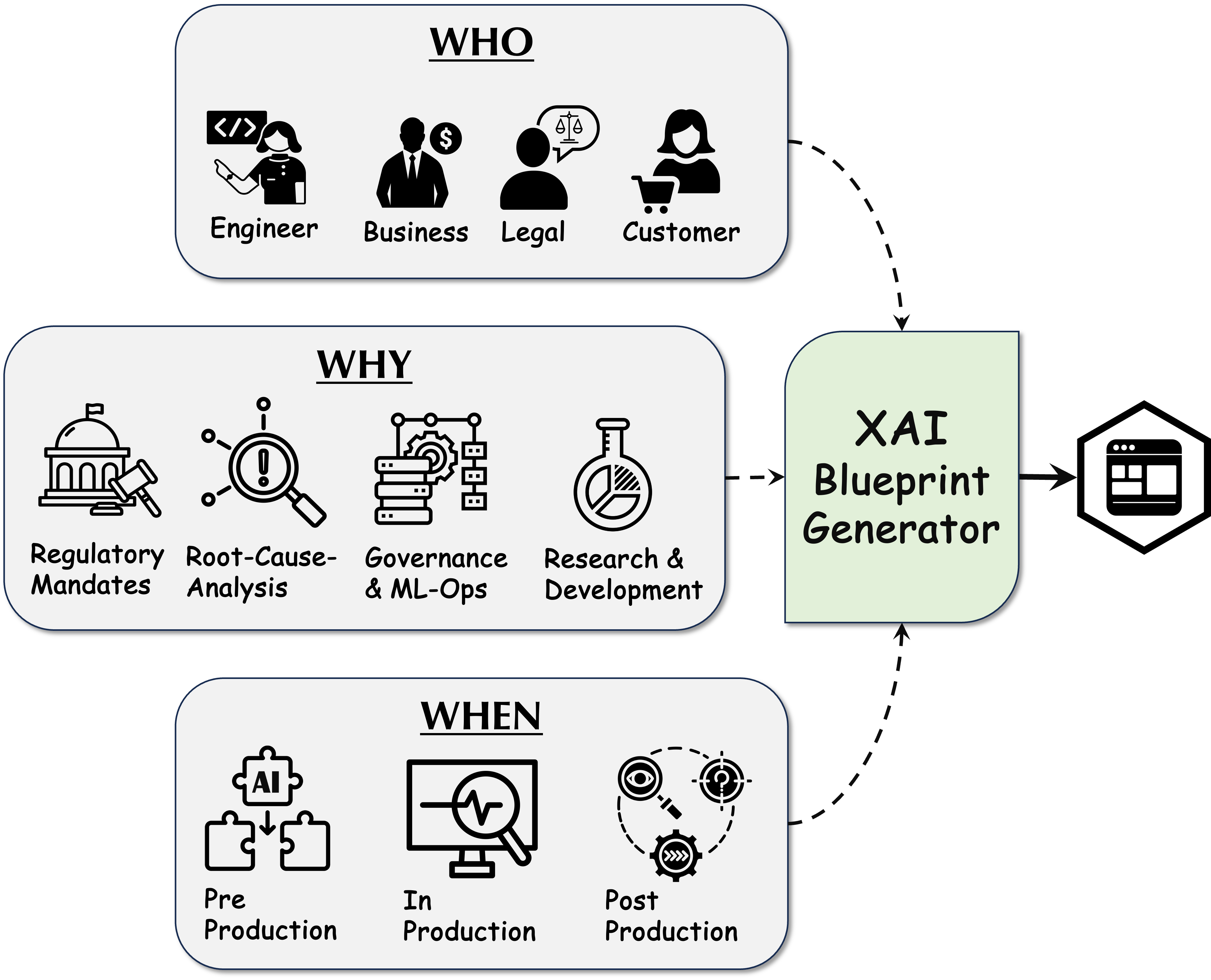}
    \caption{Our proposed framework using three main inputs to generate an `XAI Blueprint' for an organization. For an example of a XAI blue print, see \cref{tab:XAI-blueprint-gen-example} and discussions in \cref{ssec:xai-specturum}.}
    \label{fig:xai-blue-print}
\end{figure}

\subsection{Our Proposed Solution: XAI Blueprint Generation}
\label{ssec:XAI-blue-print}

Any enterprise or entity that needs XAI (as an independent or part of any TAI solution), ought to ask three questions. The sheer number of \textbf{feasible answers} to the following questions (see \cref{fig:xai-blue-print}),
\begin{enumerate}[label=\ding{43}]
    \item  \textit{\textbf{Who} needs XAI?}
    \item  \textit{\textbf{Why} is XAI needed?}
    \item  \textit{\textbf{When} is XAI needed?}
\end{enumerate}

\noindent
can easily overwhelm a medium size to large organization, and, therefore, can render the adoption of XAI impossible. 
\begin{table}[htbp]
    \centering
    \caption{Generating an `XAI Blueprint' based on our proposed framework. We propose prioritizing answers to ``Who, When, and Why XAI is needed?'' and produce a meaningful XAI Blueprint comprising questions to be addressed in the adoption phase.}
    \begin{tabular}{|p{0.1\textwidth}|p{0.12\textwidth}|p{0.23\textwidth}|p{0.55\textwidth}|}
        \hline
        \rowcolor{gray!30}        
        \textbf{Who?} & \textbf{When?} & \textbf{Why?} & \textbf{A Sample Generated XAI Blueprint} \\
        \hline
        AI Engineer & Pre-production & Help choose the best AI Model Type so that it Passes the New Regulatory Mandates by EU. & \cellcolor{gray!10}\begin{itemize}[leftmargin=*,label=\ding{43}]
            \item Does AI Model Outputs Classification or Regression?
            \item Is Uncertainty Quantification Needed?
            \item Do Legal Mandates Require Non-expert Reasoning Logic?
            \item Training Data should Exclude Sensitive Features
        \end{itemize}\\
        \hline
        Legal Team & In-production & Response to a Customer's Inquiry: `\textit{Why my Loan Application was Denied?} & \cellcolor{gray!10}\begin{itemize}[leftmargin=*,label=\ding{43}]
            \item What is the Identified Risk Level using EI-AI Act?
            \item Governance: Which AI Model was used to Make Decision for Current Applicant?
            \item If a Black-box AI Model, Estimate how much Additional Time is Required to Produce the XAI Report.
            \item What are the Mitigation Plans if AI Model shows any Legal Violations?
        \end{itemize}\\
        \hline
        Business Analyst & Post-production & RCA: Why Loan Default Rates has Increased in the Past Quarter & \cellcolor{gray!10}\begin{itemize}[leftmargin=*,label=\ding{43}]
            \item Governance: Business-friendly Features
            \item Compare against Historical Trends.
            \item If the AI Model was Updated or Retrained, which KPIs were Used to Validate the Updated Model?
            \item Examine if the AI Model is still Relevant: Has Concept or Data Drift Occurred?
        \end{itemize} \\
        \hline
    \end{tabular}
    \label{tab:XAI-blueprint-gen-example}
\end{table}

In our solution, we propose a flexible framework which answers above would be passed to an XAI Blue Generation engine to create appropriate actions, setups, and requirements to enable XAI. In \cref{tab:XAI-blueprint-gen-example}, we demonstrate this via three different end users and AI product at three different stages mentioned above. 

\section{Implementation Framework}
\label{sec:impl-framework}

In the remainder, we provide a few examples of how such frameworks are needed to implement, monitor, and enforce TAI. One point to consider is that adopting one framework is not mutually exclusive. We claim that, more often than not, depending on the context, different frameworks ought to be implemented in order to facilitate integration of `trustworthiness' in an AI-system. For example, if a private company uses an AI tool to help its employees for internal use, it may use a less stringent TAI framework\footnote{To the extent required by law.} as opposed to an AI-platform sold to external clients. Each of these entities need their own TAI framework, respectively. We remark that it is reasonable that such frameworks are different since each party has its own goal, legal requirements, or resources altogether. 

\subsection{Trustworthy-By-Design}
\label{ssec:trustworthy-by-design}

Given `trustworthiness' (or any other characteristics of TAI) is enforced before or during the designing step of product development, any attribute associated with TAI is translated as \textit{additional constraints}. 
These constraints can, in effect, restrict the choice of AI model type, training and/or testing procedure, acceptance criteria, and set of objective functions to be used by an AI model, \emph{a priori}. On the bright side, however, such seemingly stringent conditions can help build an AI product that is-- to a reasonable degree-- `\textit{trustworthy-by-design}'\footnote{It is sometimes helpful to consider example products in the \textit{physical world}. In contrast to digital products (e.g. Email), products in physical world provide a more intuitive sense of `rights', `legal mandates', `reliability', or `accountability'.}.

\begin{figure}
    \centering
    \includegraphics[width=0.9\linewidth]{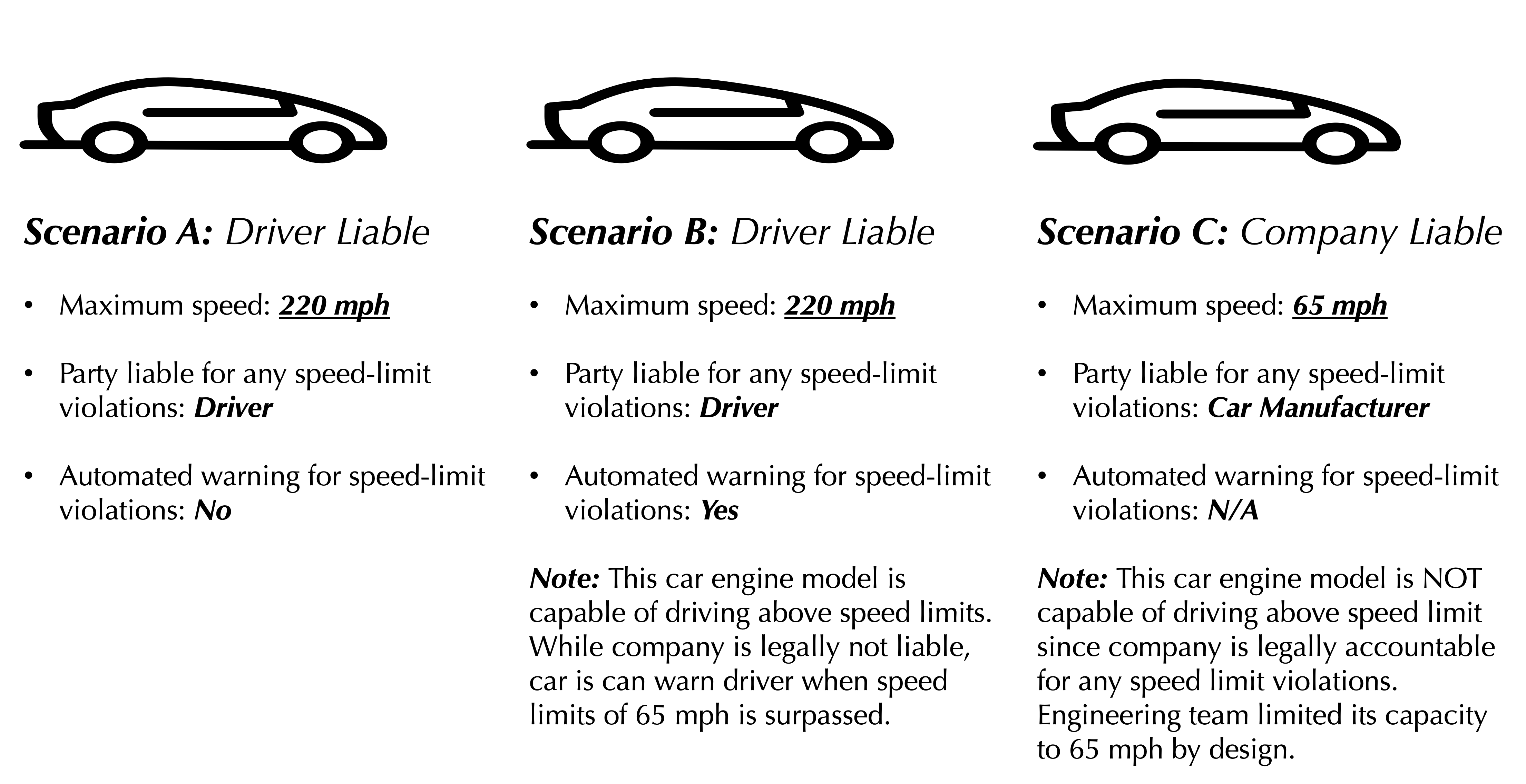}
    \caption{Illustration depicting how three hypothetical scenarios for legal liabilities can affect  the design choice and engineering of any new AI product. We are using the example of a new car being manufactured to simplify the topic of `trustworthy-by-design' for an AI system. In scenario C, the car is inherently incapable of passing the speed limit.}
    \label{fig:car-model-example-scenarios}
\end{figure}

Let us use a simple familiar example. Consider a car manufacturing company about to build and release a new model. 
For the sake of example, let's consider the following scenarios concerning legal mandates on speed-limit violations (see \cref{fig:car-model-example-scenarios}):
\begin{enumerate}[label=\ding{43}]
    \item \textbf{Scenario A:} Driver is liable for any speed-limit violations. 
    \item \textbf{Scenario B:} While driver is still liable for any speed limit violations, car companies are now required (by law) to warn drivers when surpassing the speed limit of 65 mph.
    \item \textbf{Scenario C:} New legislation holds any car company liable for any speed-limit violations. Lawmakers expect that the new car models \textit{by design} could not support speeds above 65 mph. 
\end{enumerate}

\noindent 
We hope that the example of a new car design choice by legal and engineering team states the challenges with defining a `trustworthy-by-design' framework when building an AI product. In scenario C, the engineering team designs the car engine capacity, so it is \textit{incapable} of passing a preset speed limit. 
\subsubsection{Need-to-Know-Basis}
\label{sssec:need-to-know-basis}

Despite their remarkable performance, applications built using LLMs or GPTs have shown weak to no strains against revealing `too much' information to human user (\citet{greshake2023not}). Several examples of how human users `design prompts'\footnote{Also known as `prompt injection'.} (or inputs) to extract sensitive information from AI models (\citet{liu2023jailbreaking,yao2024survey}). For example, a 13-year-old student interacting with an AI tutoring software should not be provided information on how to use drugs-- at least without the supervision of a teacher in the AI-student engagement session. With no universal solution readily available, concerns for user safety, privacy, \newabbr{IP}{Intellectual Property} theft, or fraud using `adversarial attacks' on AI models is an active research topic, cf. \citet{qiu2019review}. 

An AI-powered system should only `\textit{know}' the `\textit{knowledge}' needs to fulfill the task(s) it is built for. For example, one strategy is to ensure that training data (along with its relevant features or meta-data) that is intended to be used and build an AI product, must not contain information that is not relevant to the use-case. 
Overall, mitigation strategies in making AI products resilient against adversarial attacks and prompt injection depend on factors such as compute resources, risk level, and legal constraints, cf. \cite{rai2024guardian}.

\subsection{Trustworthy Assurance}
\label{ssec:trustworthy-by-assurance}

Unlike `trustworthy-by-design' (see section \ref{ssec:trustworthy-by-design}), `trustworthy assurance' aims to test and verify trustworthiness (or a subset of attributes of TAI) `\textit{after}' creation of any (AI) product or services. Thus, making decisions at the design and/or training step to ensure `trustworthiness' may not be required. In rare cases, an AI model can first be \textit{trained} while fully disregarding any TAI attributes. Next, `Trustworthy Assurance' team proceeds to conduct its family of predefined tests for required attributes of TAI. If any of those tests failed, results would be reflected in meaningful reports and can be handed to the engineering and legal teams for potential updates or corrections to the original model. While this approach has been adopted by many companies for the past few years, we do not recommend this as many countries are anticipated to pass laws favoring or mandating `trustworthy-by-design' methodology for implementation of TAI in the private sector. 

\subsection{Trustworthy via Continuous Monitoring and Improvement}
\label{ssec:trustworthy-by-continuous-monitoring}

As the title suggests, this framework is not exclusive from the ones discussed in \ref{ssec:trustworthy-by-design} and \ref{ssec:trustworthy-by-assurance}. Accepting `\textit{The only constant in life is change}'\footnote{The Greek philosopher Heraclitus is credited with this quote.}, any AI model/product--regardless of its initial state-- should be continuously monitored, tested, and improved. Reasons for adhering to this philosophy goes above ensuring `trustworthiness'. Continuous monitoring and improvement life-cycle has been applied to traditional software development for years (for a survey on approaches and practices, cf. \cite{shahin2017continuous}, \cite{mishra2020devops}). More recently, similar frameworks such as \newabbr{ML-Ops}{Machine Learning Operations} introduced to facilitate challenges associated with continuous development, monitoring and deployment of AI/ML models in the production settings for any business (cf., \cite{kreuzberger2023machine}, \cite{symeonidis2022mlops}, and \cite{testi2022mlops}). 
\subsection{Our Proposed Solution}
\label{ssec:our-set-formalize-measure-framework-solution}

We propose a risk-based and flexible framework to help enterprises operationalize TAI in their business. 

\begin{figure}[htpb]
    \centering
    \includegraphics[width=0.8\linewidth]{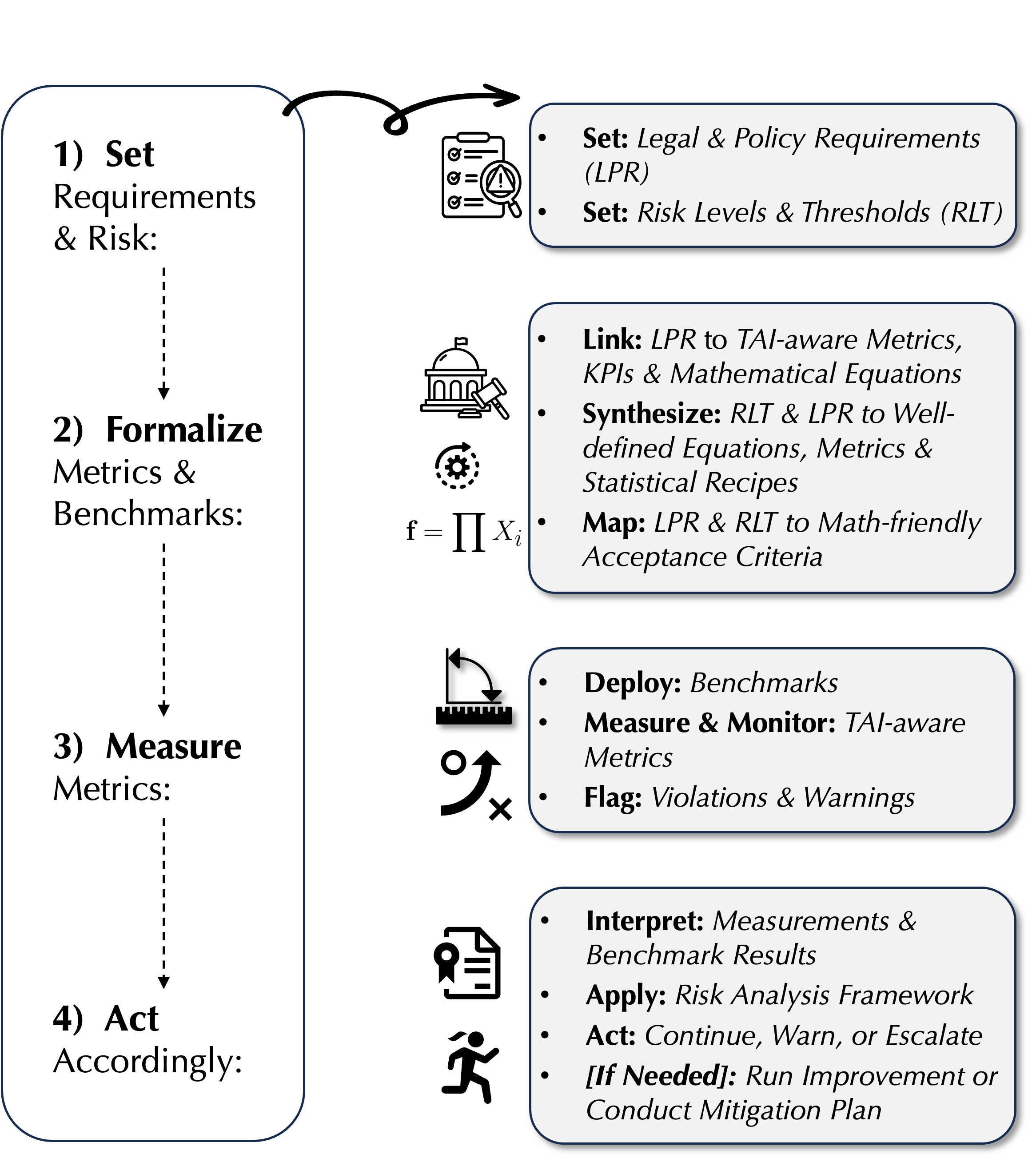}
    \caption{Our general-purpose framework, SFMA, to help any enterprise implement and operationalize TAI within their organization. For description, see discussion in \cref{ssec:our-set-formalize-measure-framework-solution}.}
    \label{fig:sfma-our-framework}
\end{figure}
\noindent 

Our proposed framework consists of four distinct steps (see \cref{fig:sfma-our-framework}):
\begin{enumerate}
    \item \textbf{Set:} Goal is to clearly set \newabbr{LPR}{Legal \& Policy Requirements} along with \newabbr{RLT}{Risk Levels and Thresholds} that have to be considered with the AI-system. 
    \item \textbf{Formalize:} LPR and RLT to define or select TAI-aware measurable metrics, KPIs and acceptance criteria. In addition, formulate proper equations to set up benchmarks. 
    \item \textbf{Measure:} Upon running benchmarks or measuring the metrics from Formalize step, record potential violation flags based on configured acceptance criteria. 
    \item \textbf{Act:} Interpret findings, metrics, and flags from previous step back to non-technical (if needed) implications. If necessary, escalate or suggest mitigation plans to minimize risk (of violation). 
\end{enumerate}

\section{A Few Suggestions for a Viable Path Forward}
\label{sec:solution}

\subsection{Continue Supporting Academic Research in Trustworthy AI}
\label{ssec:pub-tai-globally}

\newabbr{CSET}{Center for Security and Emerging Technology}\footnote{\href{https://cset.georgetown.edu/}{CSET}, based at Georgetown University, is a `\textit{think-tank}' focused on supporting decision makers using data-driven analysis.} analyzed prior scientific publications on topics related to TAI. Using a thorough and systematic analysis to contextualize the trustworthy AI terms in more than 30,000 scientific publications, CSET identified 18 clusters relevant to publications in set or subset of TAI, \cite{toney2023who}. 
\begin{table}[htbp]
    \centering
    \caption{Top publishing institutions in trustworthy AI research clusters
identified by CSET. See \cite{ETOTopPublishersAI2024} and \cite{toney2023who} for more information on how the clusters are defined.}
    \label{tab:universities-top-publishers}
\begin{threeparttable}
    \begin{tabular}{|p{0.47\textwidth}|p{0.47\textwidth}|}
        \hline
        \rowcolor{gray!30}        
        \textbf{University} & \textbf{Country} \\
        \hline
        \hline
        Arizona State University & USA \\
        Carnegie Mellon University & USA \\
        Massachusetts Institute of Technology & USA \\
        University of California Los Angeles & USA \\
        University of California Berkeley & USA \\
        University of Notre Dame & USA \\
        Google, LLC\tnote{*} & USA \\
        \hline
        Chinese Academy of Sciences & China \\
        Nanjing University & China \\
        Tsinghua University & China \\
        \hline
        Nanyang Technological University & Singapore \\
        \hline
        Darmstadt University of Applied Sciences & Germany \\
        \hdashline
        EURECOM\tnote{\dag} & France \\
        \hdashline
        University of Luxembourg & Luxembourg \\
        \hline        
        University of Technology Sydney & Australia \\
        \hline
        University of Waterloo & Canada \\
        \hline
    \end{tabular}
\begin{flushright}
\begin{tablenotes}\footnotesize
\item[*]  A for-profit entity.
\item[\dag] Graduate School and Research Center in Digital Science in Sophia, France. 
\end{tablenotes}
\end{flushright}

\end{threeparttable}
\end{table}
\noindent 

\noindent
The list of top publishers including only one private company that made it to the list (Google) is given in table \ref{tab:universities-top-publishers} (source: \cite{ETOTopPublishersAI2024}). 

\begin{figure}
    \centering
    \includegraphics[width=0.99\linewidth]{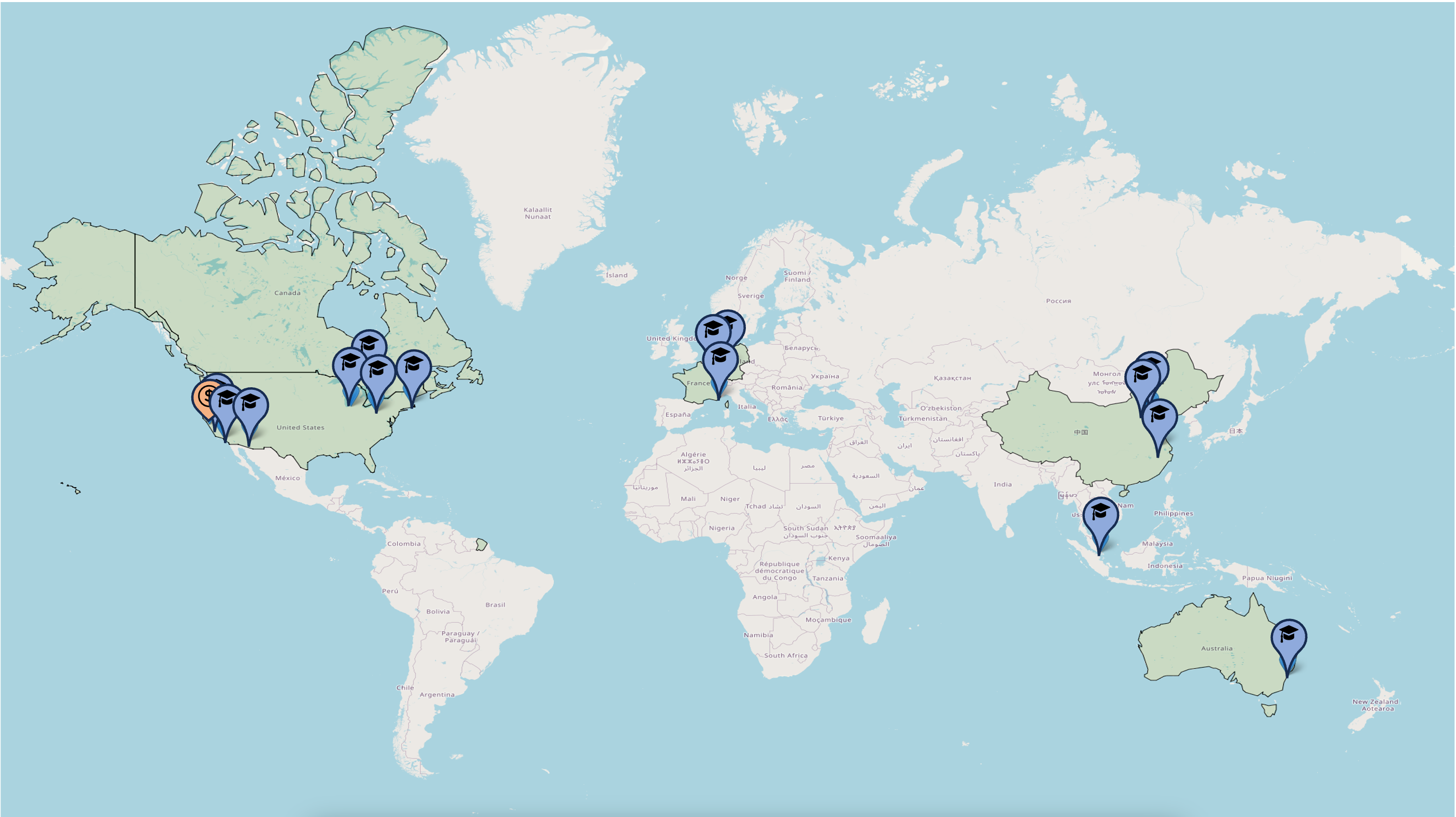}
    \caption{Top scientific publishers (universities and private research labs) in 18 topics related to TAI (see discussion in \href{ssec:pub-tai-globally}). Note that only one private entity is on this map, Google. Every other marker denotes a university. For definition of 18 topics in TAI and how the list of universities is compiled, we refer the reader to \citet{toney2023who}.}
    \label{fig:top-world-tai-publishers}
\end{figure}

\subsection{Open-Source Software (OSS): A Shiny Badge of Honor in Humans' Future History}
\topquote[James E. Rogers]{When in doubt, one can rarely go wrong by going public.}

When it comes to collaborative innovation, where we stand today-- as Isaac Newton calls it `\textit{standing on the shoulder of the giants}'-- has not always been this tangible. \newabbr{OSS}{Open Source Software} movement initiated in 1980s, currently plays a crucial and undeniable role in software and digital product life cycles. Pioneered by individual programmers, OSS ecosystem currently includes freelance developers, academia, government research facilities, for- and not-for profit companies, cf. \citet{korkmaz2024github}. The depth and breadth of OSS adoption has had a major impact on entrepreneurship, innovation, and economic growth. 
\begin{table}[htbp]
\centering
\caption{A few open-source projects that had remarkable impact on a global scale. Note that the total number of contributors (the second column) for every project is extracted from its official GitHub webpage. Data extracted on February 26, 2024.)}
\label{tabl:successful-github-repos}
\begin{threeparttable}
\resizebox{\textwidth}{!}{ 
\begin{tabular}{llll}

\toprule
Name                                                 & Number of Contributors\tnote{*} & Domain  & Release Date\tnote{**} \\
\midrule
\href{https://github.com/ansible/ansible}{Ansible}   & 5,554   & IT Automation System  & 2012  \\
\href{https://github.com/twbs/bootstrap}{Bootstrap}  & 1,387   & Front-end Development  & 2011\tnote{\dag} \\
\href{https://github.com/kubernetes/kubernetes}{Kubernetes} & 3,584 & Digital Product Deployment System & 2014\tnote{\dag} \\
\href{https://github.com/torvalds/linux}{Linux Kernel}    & 15,510  & Operating System  & 1991     \\
\href{https://github.com/opencv/opencv}{OpenCV}          & 1,564    & Computer Vision  & 2002\tnote{\dag}      \\
\href{https://github.com/openssl/openssl}{OpenSSL}         & 875    & Cryptography \& Network  & 1998  \\
\href{https://github.com/python/cpython}{Python}          & 2,606   & Programming Language & 1991      \\
\href{https://github.com/scikit-learn/scikit-learn}{Scikit-Learn}    & 2,815 & Machine Learning Library & 2007 \\
\bottomrule
\end{tabular}
} 
\begin{flushright}
\begin{tablenotes}\footnotesize
\item[*] Reported on GitHub.
\item[**] Date denotes the first time that any package was shared as OSS.
\item[\dag] Initially launched as either a commercial or an internal-use-only software.
\end{tablenotes}
\end{flushright}

\end{threeparttable}

\end{table}

\noindent 
For example, \newabbr{LOS}{Linux Operation System} is running on all `Top 500' supercomputers\footnote{Based on \href{https://www.top500.org/}{TOP500}'s (November 2023) report (\citet{top500}). Initiated in 1993, \href{https://www.top500.org/}{TOP500} is a non-profit project which publishes detailed reports and benchmark results of the 500 most powerful supercomputers in the world, biannually.}, and 96\% of top web-servers\footnote{Top one-million web-servers.}. But the best part is: Linux Kernel made it to Mars.

\subsubsection{Linux Operating System `Flying' on Mars}
Ingenuity helicopter-- nicknamed Ginny-- currently on \textbf{Mars} just completed its 72nd and final flight, \cite{NASAHelicopter2024}). 
Running on Linux\footnote{JPL used \textit{Linaro 3.4.0}-- a Linux distribution that supports Qualcomm Snapdragon processors-- in Ingenuity helicopter.} \newabbr{OS}{Operating System}-- a fully open-source software-- Ginny's huge success in exploring Mars is hailed by many advocates of free software systems. What is remarkable is that Ginny was built collaboratively by NASA's \newabbr{JPL}{Jet Propulsion Lab}) and utilized several open-source software during different phases. Here is what Timothy Canham-- the operations lead and former software lead of the Mars helicopter project at JPL-- has to say about the role of open-source in the success of Ginny: 

\begin{quote}
``\textit{This the first time we'll be flying Linux on Mars. We're actually running on a Linux operating system. The software framework that we're using we open-sourced it a few years ago. So, you can get the software framework that's flying on the Mars helicopter, and use it on your own project. It's kind of an \textbf{open-source victory}, because we're flying an open-source operating system and an open-source flight software framework and flying commercial parts that you can buy off the shelf if you wanted to do this yourself someday.}''\cite{IEEEMarsHelicopter2024}
\end{quote}

\begin{figure}
    \centering
    \includegraphics[width=0.80\linewidth]{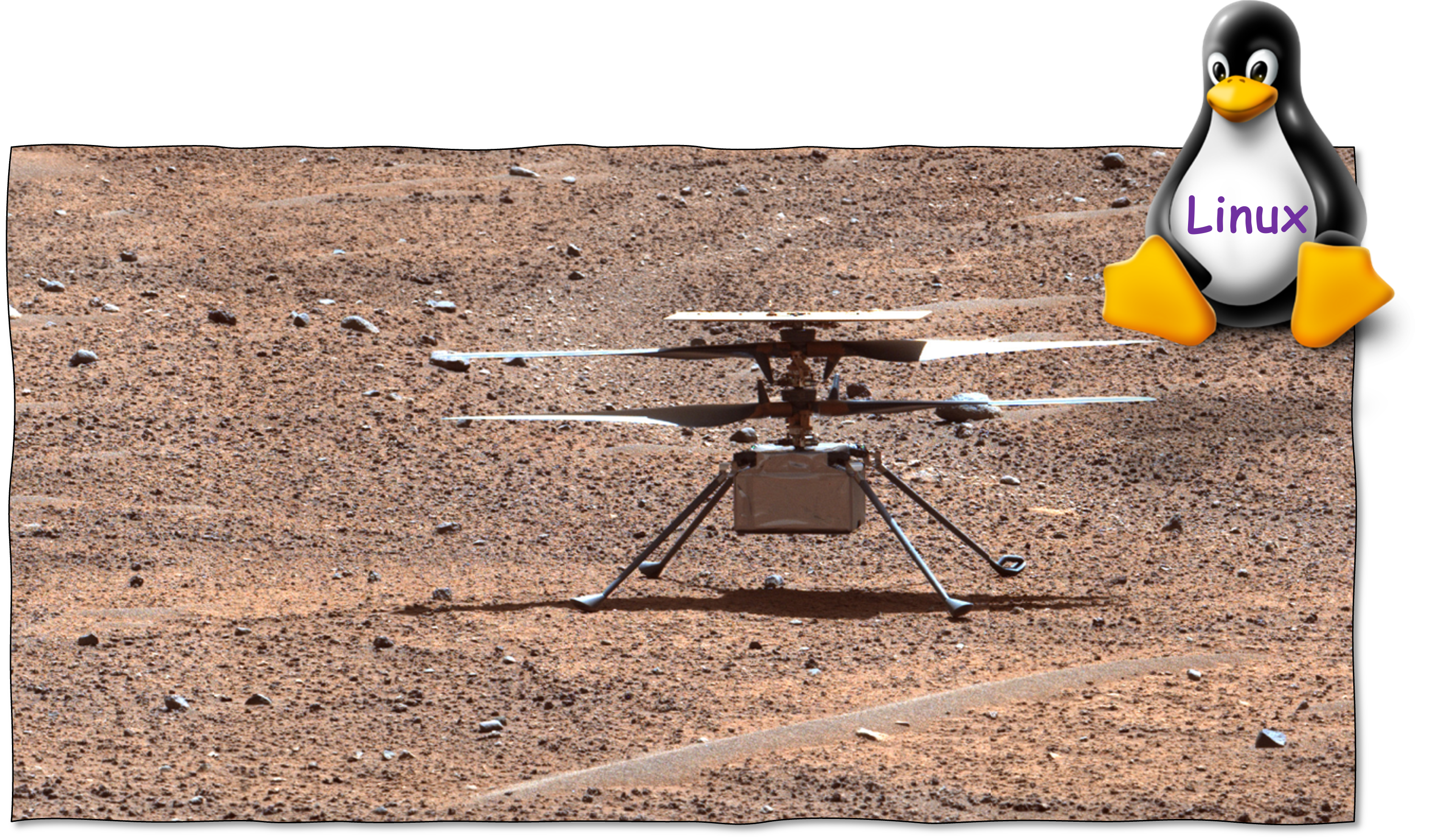}
    \caption{Ingenuity helicopter (nicknamed \textit{Ginny}) was photographed on the surface of Mars on August 2nd, 2023, by another current `resident' of Mars-- the Perseverance Mars Rover. The software framework used in Ginny by NASA and JPL was based on Linux kernel-- an open-source software. Credit: NASA/JPL-Caltech-ASU/MSSS.}
    \label{fig:linux-mars-ginny}
\end{figure}
\noindent
\subsubsection{Let's not Take Open-source for Granted: Hiding Scientific Discoveries for `\textit{Job Security}' in the Past}

Making it this far to this level of openness in sharing knowledge and collaborations amongst fans, enthusiasts, and academics has not been always this easy. Why? Here is a good example from the history of mathematicians in Europe. `\textit{Scipione del Ferro}', a Renaissance mathematician from Italy is credited with the first to discover an analytical solution for a subset of cubic equations of the form $x^3 + ax = b$. His ingenious solution approach led to further innovations in mathematics and complex numbers. Yet, Scipione kept this achievement secret until his deathbed, when he finally shared his notebook with his student, `\textit{Antonio Maria Fiore}', cf. \cite{feldmann1961cardano} and \cite{britannicaFerro}. 

This level of secrecy was common during the Renaissance era, as scholars feared competition and the potential loss of their academic positions, leading many to withhold their most \textbf{significant discoveries} as a form of ``\textbf{job security}''.

\subsection{Open-sourcing AI: \textit{Free-as-in-Beer} \emph{vs} \textit{Free-as-in-Speech}}
\topquote[Jon I. Lovett]{When in doubt, mock the powerful, not the powerless.}

The choice of what and how `freely' AI should be shared and available online would be shared has roots in now famous philosophy by \newabbr{GNU}{GNU's Not Unix} which articulates the term `free software':
\begin{quote}
``\textit{Free software means software that respects users' freedom and community. Roughly, it means that the users have the freedom to run, copy, distribute, study, change and improve the software. Thus, ``free software'' is a matter of liberty, not price. To understand the concept, you should think of ``free'' as in ``free speech'', not as in ``free beer''. We sometimes call it ``libre software'', borrowing the French or Spanish word for ``free'' as in freedom, to show we do not mean the software is \textit{gratis}.}''.\cite{GNUFreeSoftware2024}
\end{quote}

\noindent 
The movement behind `Open-Sourcing AI' has gained momentum, mainly, thanks to existing and widely-accepted OSS ecosystem and community culture. However, inherent to the nature of AI products, existing open-sourcing software frameworks cannot capture the entire complexity associated with open-sourcing AI products. In other words, while sharing an algorithm that were utilized to build and train an AI model (along with the actual computer code in human readable form) is a good start, we argue that `true' AI open-source requires additional components and considerations. The list below is a good example when major players consider open-source AI frameworks.

\begin{enumerate}[label=\ding{43}]
    \item \textbf{A Trained AI Model:} This can vary, as an AI model could be shared as an executable, binary, or byte-code, e.g. weights and biases of in a DNN. 
    \item \textbf{(Or) Recipe to the Train AI Model}: Alternatively, an AI model's architecture, e.g. in a DNN, number of Neurons per layer, activation function type, can be shared. Yet, the actual values of weights and biases should be (re-)produced using a `training recipe'.  
    \item \textbf{Data:} Governance, license, meta-data, or recipes to generate (synthetic or generic) data
    \item \textbf{Deployment:} Includes runtime dependencies, libraries, OS and system settings, etc. A common popular solution is the Docker-Containers.
    \item \textbf{Provenance:} Providing the origin of data (training or test data) along with algorithm, recipe pipeline(s), or additions.
    \item \textbf{Legal:}  With the highly-anticipated regulatory provisions to be enforced on AI, we argue that successful platforms such as GitHub can take the lead in educating their users on geographically-varying legal ramifications of their high-stakes AI projects. In short, licensing of any new (free) AI project may not be as easy as adding a simple \textbf{LICENSE.md} or \textbf{README.md} to the shared repository. 
\end{enumerate}

\subsection{Where is AI Headed: A Few Insights from GitHub Trends}
\label{ssec:where-is-ai-header-github}

In its 2023 Octoverse\footnote{According to the official GitHub website, `\textit{Octoverse}' is an annual report sharing the state of open source by reporting data-driven insights and activity data collected from GitHub platform. For details on the methodology, we refer the reader to \citet{dohmke2023sea}.}, using user-activity data, GitHub summarizes trends driving OSS globally. We invite reader to read this report here, \citet{githubblog2023}. Below, we highlight items driven by the Generative AI movement:

\begin{enumerate}
    \item \textbf{Leader board:} USA, India, and Japan are leaders on the scoreboard based on the total number of individual contributors to Generative AI projects.
    \item \textbf{India to dethrone USA:} Currently ranked as second, India is projected to dethrone USA as the largest developer community on GitHub by 2027. A Major driver has been the large scale use of open banking system and government welfare system\footnote{Other notable examples are Mercado Libre, Latin America's largest e-commerce ecosystem, and Pix, Brazil's real-time payment infrastructure. Marcado Libre used GitHub to automate deployment and tests to aid its developers. The Central Bank of Brazil recently made Pix's communication protocols open-source.}
    \item \textbf{AI-powered Co-pilots:} GitHub developers are actively \textbf{experimenting} and \textbf{building} their projects using AI-powered tools, e.g. code co-pilot. 
    \item \textbf{Paradigm shift in experimenting with AI:} AI developer and experimenters are shifting from more ``traditional'' libraries such as \textit{TensorFlow} and \textit{PyTorch} to pretrained and foundational models, LLMs, and even ChatGPT API.
    \item \textbf{Generative AI amongst most popular projects:} For the first time, in 2023, open sources Generative AI repositories made it to the top 10 most popular projects \footnote{Ranked by `contributor count'.}. Also, year 2023 saw the largest number of first-time contributors to OSS projects; 2.2M (120\% increase from 2017). 
    \item \textbf{Building Cloud-ready Products}. Developers are scaling cloud-native applications using declarative languages and Git-based \newabbr{IaC}{Infrastructure-as-Code} workflows. Standardization in cloud deployments, notable surge in using Dockers and containers, and other cloud-native technologies, signaling a shift towards product-ready mentality.
    \item \textbf{Open-source AI Innovation is Healthy:} Consider the top 20 open source AI projects on GitHub in 2023: They are diverse in nature, application, and ownership (individuals or private companies). Some of the most popular AI projects have been developed and maintained by individual SMEs with no affiliations to for-profit companies. This is another indicator that open source projects in the field of AI can substantially contribute to the growth and mitigating the challenges with implementation of TAI.
\end{enumerate}

\section{Summary and Next Steps}
\label{sec:summary}

We reviewed definitions of TAI (and its ``synonyms'') shared by entities such as UNESCO, IEEE, and NIST to consolidate the most frequently mentioned `attributes' of TAI. 
We suggest `comprehending' the complex topic of TAI through the lens of characterizing its attributes or intrinsic properties. In the past decade, there has been multi-disciplinary research to project concepts such as `fairness', `biased outcomes', `risk and security', `transparent', etc onto AI research. Yet, inherent complexity and subjective nature of aforementioned topics do not render a concrete \mbox{`\textbf{\textit{one-size-fits-all}}'} TAI framework. To make matters more challenging, countries, their legislative entities, and international organizations have taken philosophically distinct path forward to regulate AI and standardize TAI.

Here, we are offering a multi-prong path forward. Our main message is to first and foremost, empower the open source movement. If history has shown us, panic is not the best guidance. We strongly advise against over-regulation which could hinder innovation and growth of the open source community. To support our claims, we have reported exciting recent trends in AI derived from user activity data shared by GitHub in 2023. We do not underestimate the potential risks of modern AI model such as the GPT family and LLMs. Yet, to mitigate potential risks, supporting academic research and enabling open-source communities to access AI models and compute platforms are very critical factors. 

We demonstrate how existing frameworks such as the Rumsfeld Risk Matrix (RMM) can be applied to enable AI engineers plan for risks associated with the behavior of their AI systems. As EU-AI-Act has approached AI regulation through `risk framework', it is imperative to combine existing ERM and any AI-system and its `uncertainty level'. Having proper mapping between risk level (based on of the type of uncertainty type and its severeness) would be crucial for companies to avoid hefty fines provisioned in EU-AI-Act. 

We proceed to introduce our (meta-)framework, `\textbf{Set\textrightarrow Formalize\textrightarrow Measure\textrightarrow Act}' to adopt and implement TAI. This is an example framework designed to enable various personas and decision makers involved in AI-product life-cycle. By nature, we aimed to have this (meta) framework generic enough so it can serve different for- or non-profit entities at various AI-product adoption level. 

We hope this series can trigger enthusiasm in SMEs across different domains. We firmly believe that building efficient and helpful TAI frameworks requires an open and collaborative task. Areas such as TAI in judiciary systems, education, national or international security, or healthcare are only a few examples. The scale and complexities within these domains demand honest and multi-disciplinary collaboration. In part two of this series, we aim to provide more technical, statistical, and algorithmic details around TAI framework with focus on identifying proper metrics.

\clearpage
\newpage
\section{About the Authors}
\noindent 
\begin{multicols}{2}
\begin{figure}[H]
\includegraphics[width=0.65\linewidth]{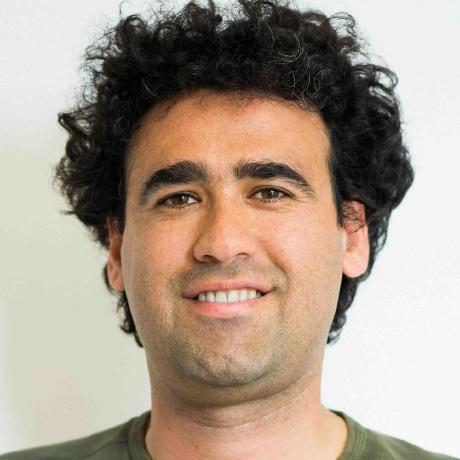}
\nonumber
\end{figure}

\noindent 
\textbf{Mohamad M. Nasr-Azadani} \\
Founder \& Chief Executive Officer\\
\href{mailto:mohamad@impartialZ.com}{mohamad@impartialZ.com}\\
\textit{Impartial GradientZ}\\ \\
\textbf{About:} Mohamad is the founder of \href{https://impartial-gradientz.com/}{Impartial GradientZ}-- an advisory firm helping clients with bridging the gap between cutting-edge R\&D in AI and Machine Learning and complex industry challenges. In the past, Mohamad served as a Principal Data Scientist at \href{https://www.accenture.com/us-en}{Accenture} leading applied R\&D in AI-- ML-Ops, Differentiated Hardware and AI Model Deployment, Causal Inference, Physics-informed AI, and Digital Twins. He holds numerous patents, peer-reviewed \href{https://scholar.google.com/citations?user=alBGGikAAAAJ&hl=en&oi=ao}{publications}, and presented at international conferences. 
Prior to Accenture, Mohamad received his Ph.D. in Mechanical Engineering from the \href{https://www.ucsb.edu/}{ University of California at Santa Barbara}. His research was focused on `Computational Fluid Dynamics' (CFD), and developed computational \& mathematical models of `\href{https://youtu.be/UR5NoaqpCok?si=h155weqpOaUPlo0-}{underwater avalanches}' occurring in the ocean. Mohamad loves to learn new topics, help startups grow, and build AI models from scratch. 
\columnbreak
\begin{figure}[H]
\includegraphics[width=0.65\linewidth]{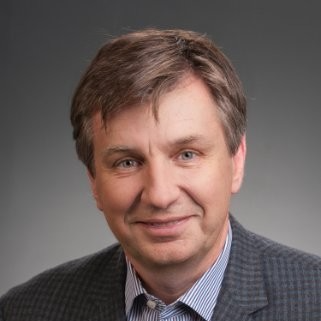}
\nonumber
\end{figure}

\noindent 
\textbf{Jean-Luc Chatelain} \\
Founder \& Senior Managing Director\\
\href{mailto:jlc@veraxcap.com}{jlc@veraxcap.com}\\
\textit{Verax Capital Advisors}\\ \\
\textbf{About:} Jean-Luc is the founder of \href{https://veraxcap.com/}{Verax Capital Advisors}, offering strategic guidance in AI and digital transformation. In the past, Jean-Luc served as the Global CTO for Applied Intelligence at \href{https://www.accenture.com/us-en}{Accenture}, leading AI initiatives with a global team of 25,000 focused on AI. Jean-Luc also acted as a trusted advisor for Accenture's global 2000 clients, guiding their AI and analytics modernization journeys, and held senior executive roles at \href{https://www.ddn.com/}{DDN} and \href{https://www.hp.com/us-en/}{Hewlett Packard}. He holds engineering and technology degrees from institutions in France, and throughout his career has created and co-invented patented technologies related to data protection, code certification, instrumentation planning, and artificial intelligence. Currently, Jean-Luc serves on advisory boards such \href{https://sambanova.ai}{SambaNova.ai}, \href{https://writer.com/}{Writer.ai}, \href{https://www.hyperscience.com/}{Hyperscience}, \href{https://ligadata.com/}{LigaData} and venture capital firm \href{https://www.forteventures.com/}{Forté Ventures}. Jean-Luc is currently a member of the board of directors at \href{https://www.k2view.com/}{K2view}. 

\end{multicols}

\clearpage
\newpage
\appendix
\section{Appendix}
\label{sec:Appendix}

\subsection{Nomenclature}
\printnomenclature[1cm]


\subsection{Guiding Principles for Trustworthy AI Released by Various Entities}
\label{ssec:apx-guiding-principles-from-entities}

\subsubsection{NIST: Characteristics of a Trustworthy AI System}
\label{sssec:NIST-definition-trust-ai}
An agency of the United States Department of Commerce, NIST's mission is to enable innovation and competitiveness in the American industries, cf. \citet{NISTAbout}. In its first published draft, NIST approaches realization of TAI through a \newabbr{RMF}{Risk Management Framework}, i.e. `AI-RMF 1.0'. The following characteristics constitute the foundation for an AI system to be considered `\textit{trustworthy}'(\citet{NISTAI}):
\begin{enumerate}
    \item Valid and Reliable
    \item Safe 
    \item Secure and Resilient
    \item Accountable and Transparent
    \item Explainable and Interpretable
    \item Privacy-enhanced
    \item Fair-- with Harmful Bias Managed
\end{enumerate}

\subsubsection{UNESCO: Ten Principles to Achieve Ethical AI}
\label{ssec:UNESCO}

UNESCO held its 41st session in November 2021, Paris. With more than 190 members, UNESCO laid out \textbf{ten principals} to guide countries and private entities for the development of `Ethical AI' with a focus on human-rights centered approach (see \citet{unesco2021recommendation}):

\begin{enumerate}
    \item Proportionality and `\textit{Do No Harm}'
    \item Safety and Security
    \item Right to Privacy and Data Protection
    \item Multi-stakeholder and Adaptive Governance \& Collaboration
    \item Responsibility and Accountability
    \item Transparency and Explainability
    \item Human Oversight and Determination
    \item Sustainability
    \item Awareness \& Literacy
    \item Fairness and Non-discrimination
\end{enumerate}

\subsubsection{IEEE: `Ethically Aligned Design' of Autonomous \& Intelligent Systems}
\label{sssec:IEEE-EAD}
IEEE\footnote{IEEE is the world's largest technical \& non-profit association of professionals and engineers whose mission has been ``\textit{dedicated to advancing technology for the benefit of humanity}''\citet{IEEEmission}}-- one of the most prominent global societies of engineers and technical professionals-- in 2016 released \newabbr{EAD}{Ethically Aligned Design} principles (\citet{ieee2017ethicallyaligned}) to recommend standards in areas pertaining to \newabbr{A/IS}{Autonomous \& Intelligent Systems}. There has been a second version (EADv2) according to IEEE's announcement: 
\begin{quote}
``\textit{The most comprehensive, crowd-sourced global treatise regarding the ethics of Autonomous and Intelligent Systems available today, EADv2 provides an open platform for thought leadership and action to prioritize value-driven, ethically-aligned design for autonomous and intelligent systems.}'' \citet{IEEEEADv2News}
\end{quote}

\noindent
In its mission, IEEE shares the following general principles aimed to realize `ethical' A/IS systems:
\begin{enumerate}[label=\textbf{Principle \arabic*:}, leftmargin=*, align=right]
    \item Human Rights
    \item Prioritizing Well-being (of humans)
    \item Accountability
    \item Transparency
    \item A/IS Technology Misuse and Awareness of it
\end{enumerate}

\noindent
What is imperative about the above principles is that the `IEEE Global Initiative on Ethics of Autonomous and Intelligent Systems' is motivated to address these concerns pragmatically via a `\textit{solutions-by-design}' approach\footnote{Another common approach is `solutions-by-review' which attempts to monitor and fix the ethical issues after the product is created, i.e. \emph{post hoc} enforcement.}. 

\subsubsection{OECD: AI Principles and Recommendations for Policy Makers}
\label{sssec:OECD}
OECD is an international organization with 38 member countries. Primary mission of OECD is to stimulate economic growth and improve world trade. In doing so, OECD commits to honor democracy and market economy. 
\begin{quote}
``\textit{The OECD AI Principles promote use of AI that is \textbf{innovative} and \textbf{trustworthy} and that respects \textbf{human rights} and \textbf{democratic values}. Adopted in May 2019, they set standards for AI that are practical and flexible enough to \textbf{stand the test of time}.}''(\citet{OECDAIPrinciples})
\end{quote}

\noindent
Here are the five complementary values-based principles shared by OECD and have been endorsed by 46 countries (\citet{OECDAIPrinciples}):
\begin{enumerate}[label=\textbf{Principle \arabic*:}, leftmargin=*, align=right]
    \item Inclusive growth, sustainable development, and well-being
    \item Human-centered values and fairness
    \item Transparency and explainability
    \item Robustness, security, and safety
    \item Accountability
\end{enumerate}

\subsection{Example Product Requirement Document: To Build and Deploy a Trustworthy AI System for Credit Risk Score Assessment}
\label{ssec:apx-template-requirements}

\textbf{Product Requirements Document: Creditworthiness Risk Assessment}
\begin{enumerate}[label=\ding{114}]
\item \textbf{Introduction:}
\begin{enumerate}
    \item \textbf{Purpose:} Estimate `creditworthiness risk score' of individuals using AI
    \item \textbf{Scope:} Available to our clients in 42 countries in North America, Europe, and Africa
\end{enumerate}

\item \textbf{Primary Objectives:}
\begin{enumerate} 
    \item Improve loan decision accuracy
    \item Ensure geographically-dependent legal compliance
    \item Reduce loan application processing time
\end{enumerate}

\item \textbf{Stakeholders:}
\begin{enumerate}
    \item Loan originators
    \item Legal and compliance team
    \item IT development team
    \item Data Science team
    \item \textit{[External]} Loan applicants
\end{enumerate}

\item \textbf{Product Features:}
\begin{enumerate}
    \item Credit Risk Score Estimator:
\begin{itemize}
    \item \textbf{Inputs:} Variable. Potential features: income, debt-to-income ratio, credit history, employment, assets, education level, demographics (only in countries where legally permitted) \dots
    \item \textbf{Output:} A numerical creditworthiness risk score with clear thresholds defining: ``\textit{approval}'', ``\textit{rejection}'', and ``\textit{needs review}'' ranges.
    \item \textbf{AI Model:} Recommended model types are: a) Logistic Regression, b) Decision Trees, or c) Gaussian Naive Bayes. \textit{Note: Model type directly affects the level of details in the explainability feature. Consult legal team for further information.}
\end{itemize}
    \item Regulatory Compliance Module:
    \begin{itemize}
    \item Geo-Location identification: How applicant's location (country, potentially state/region) was utilized-- if any?
    \item Permissible data features. Note: These features are country and/or regional dependent. 
    \item Bias and fairness monitoring procedures. Note: Relevant metrics chosen to quantify fairness/bias and formulated recipes must be reported clearly. Exception: To comply with EU-AI-Act v0.1-2023, use EU-Fairness-Wizard internal tool.
    \item Model decision explainability features:
        \begin{itemize}
            \item \textbf{Global Explanations:} Feature importance; Parity statistics; Human-comprehensible factors influencing the risk score; Uncertainty estimates
            \item \textbf{Local Explanations (where required):} Provides applicant-specific reasons for their score/decision.
        \end{itemize}
    \end{itemize}
\end{enumerate}

\item \textbf{Reporting \& Monitoring:}
\begin{itemize}
    \item \textbf{Legal Compliance Dashboard:} Tracks key metrics across sensitive legal mandates, e.g. fairness test results across demographics as required by local laws.
    \item \textbf{AI Model Performance Tracking:} Monitors accuracy, drift, and any disparate outcomes.
    \item \textbf{Audit Logs:} Tracks all model usage, including inputs, outputs, and any regulatory disclosures.
\end{itemize}

\item \textbf{Technical Considerations:}
\begin{itemize} 
\item \textbf{Integration:} Interfaces with existing loan application products. 
\item \textbf{Security:} Adheres to the company's strict data privacy and security protocols.
\item \textbf{Scalability:} Accommodates the expected growth.
\end{itemize}

\item \textbf{Open Issues \& Constraints:}
\begin{itemize}
    \item \textbf{Legal Review:} Continuous legal counsel is needed to keep regulatory rule sets updated. Periodic external audit of AI model is highly recommended.
    \item \textbf{Data Availability:} Sourcing reliable data in some jurisdictions may be a challenge.
    \item \textbf{Explainability vs. Model Performance:} Creating explainable models might involve trade-offs with potential accuracy.
\end{itemize}

\end{enumerate}

\bibliographystyle{plainnat}
\bibliography{Zreferences}

\end{document}